\documentstyle[psfig]{aa}
\begin{document}

\thesaurus{8(13.09.6, 08.19.3, 08.13.2, 08.02.4, 08.03.4, 08.09.2 AFGL~4106)}
\title{The composition and nature of the dust shell surrounding the binary
AFGL~4106\thanks{Based on 
observations with ISO, an ESA project with instruments
funded by ESA Member States (especially the PI countries: France,
Germany, the Netherlands and the United Kingdom) and with the
participation of ISAS and NASA} \thanks{Based on observations collected at the 
European Southern Observatory, La Silla, Chile}}
\author{F.J. Molster\inst{1}, 
L.B.F.M. Waters\inst{1,2}, 
N. R. Trams\inst{3},
H. van Winckel\inst{4}, 
L. Decin\inst{4},
J. Th. van Loon\inst{1}, 
C. J\"ager\inst{5},
Th. Henning\inst{5},
H.-U. K\"{a}ufl\inst{6},
A. de Koter\inst{1},
J. Bouwman\inst{1}
}
\offprints{F.J. Molster: frankm@astro.uva.nl}
\institute{
Astronomical Institute 'Anton Pannekoek', University of Amsterdam,
Kruislaan 403, 1098 SJ Amsterdam, The Netherlands
\and
SRON Space Research Laboratory, P.O. Box 800, NL-9700 AV Groningen, The Netherlands
\and
Astrophysics Division, Space Science Department of ESA, ESTEC,
PO Box 299, 2200 AG Noordwijk, The Netherlands.
\and
Instituut voor Sterrenkunde, K.U. Leuven, Celestijnenlaan 200B,
3001 Heverlee, Belgium
\and
Astrophysical Institute and University Observatory (AIU), 
Schillerg\"a\ss chen 2-3, D-07745 Jena, Germany
\and
European Southern Observatory,Karl-Schwarzschild-Stra\ss e 2, D-85748 Garching bei M\"unchen, Germany
}

\date{received 23 februari 1999, accepted 16 June 1999}

\maketitle
\markboth{F.J. Molster et al.}{AFGL 4106}
%\markleft{F.J. Molster et al.}
%\markright{AFGL~4106}

\begin{abstract}

We present infrared spectroscopy and imaging of AFGL~4106.
The 2.4-5 $\mu$m ISO-SWS spectrum reveals the presence of a cool, luminous star
(T$_{\mbox{eff}} \approx 3750$~K) in addition to an almost equally luminous 
F star (T$_{\mbox{eff}} \approx 7250$~K).
The 5 -- 195 $\mu$m SWS and LWS spectra are dominated by strong emission
from circumstellar dust. We find that the dust consists of amorphous
silicates, with a minor but significant contribution from crystalline 
silicates.
The amorphous silicates consist of Fe-rich olivines. The presence of 
amorphous pyroxenes cannot be excluded but if present they contain much less
Fe than the amorphous olivines. 
Comparison with laboratory data shows that the pure Mg-end members of
the crystalline olivine and pyroxene solid solution series are present.
In addition, we find strong evidence for simple oxides 
(FeO and Al$_2$O$_3$) as well as crystalline H$_2$O ice.
Several narrow emission features remain unidentified.

Modelling of the dust emission 
using a dust radiation transfer code shows that large grains ($\approx 1 \mu$m)
must be present and that the abundance of the crystalline silicates is between 
7 and 15\% of the total dust mass, depending on the assumed enstatite to 
forsterite ratio, which is estimated to be between 1 and 3.
The amorphous and crystalline dust components in the shell 
do not have the same temperature,
implying that the different dust species are not thermally coupled.
We find a dust mass of $\approx 3.9 \cdot 10^{-2}$~M$_\odot$ 
expelled over a period of $\approx 4 \cdot 10^{3}$~years for
a distance of 3.3~kpc. The F-star in the AFGL~4106 binary is likely a 
post-red-supergiant in transition to a blue supergiant or WR phase.

\end{abstract}

\section{Introduction}

Late stages of evolution of both low and high-mass stars are
characterized by high mass-loss in a slow and dusty outflow.
Low- and intermediate-mass stars (up to 8 M$_\odot$) will lose a significant
amount of their initial mass on the Asymptotic Giant Branch (AGB).
More massive stars will lose 
their hydrogen-rich envelopes during a red supergiant (RSG) phase. 
The mass-loss rates can be as high as 10$^{-4}$~M$_{\odot}$/yr for the 
low-mass stars and up to 10$^{-3}$~M$_{\odot}$/yr for the high-mass stars.
Low and intermediate mass objects
evolve to the blue part of the HR diagram after the entire H-rich
envelope has been exhausted, and a planetary nebula (PN) is formed. High
mass stars either explode as a supernova while they are red supergiants,
or they evolve to the blue part of the HR diagram to become a population
I Wolf-Rayet (WR) star before ending their live.

The evolution of transition objects, i.e. stars that are in a rapid
phase of evolution between the red and blue part of the HR diagram
is not very well known. It is clear however, that mass loss plays an 
important role in determining evolutionary timescales.
Remnants of previous mass-loss phases are still in the vicinity of the
star in the form of detached dust shells. Infrared observations clearly reveal
the existence of these dust shells and  the study of the 
Circum Stellar Environment (CSE) provides essential information about 
mass-loss history and consequents of evolution.

The infrared-bright object AFGL~4106 (= IRAS~10215-5916) 
was classified as a transition object by Hrivnak et al. (1989, hereafter HKV)
based on its IRAS colours. 
Some confusion exists in the literature about its identification.
HKV quote Bidelman, who claim that IRAS~10215-5916 is
equal to CoD~-58$^{\circ}$3221. Apparently, this has sometimes
been mistaken in the literature as CPD~-58$^{\circ}$3221 and 
therefore HD~303822 (e.g. in the Tycho catalogue). 
The correct identifications should be 
HD~302821 and CPD~-58$^{\circ}$2154.

The central star was classified as a G2 supergiant by
Garc\'{\i}a-Lario et al. (1994). The combination of a cool central star
with low surface gravity and a relatively warm (i.e. young) detached
dust shell is typical for transition objects.
Garc\'{\i}a-Lario et al. (1994) found double-peaked [N {\small II}] lines in 
the optical spectrum, which they thought to originate from
an expanding envelope. 
The presence of the [N {\small II}] is surprising given the spectral type of
this object.
From the separation between the two peaks they found an expansion velocity of 
v$_{\rm exp} = 17 \pm 2$ km s$^{-1}$, which is quite reasonable for an 
AGB-wind.
The expansion velocity of the detached shell was also measured in the 
CO (J$=1-0$) rotational line and was found to be
about 40 km~s$^{-1}$ (Josselin et al. 1998). 
This is high for a low mass AGB star and suggests a massive progenitor. 
The difference with the [N {\small II}] line is remarkable and 
is probably related to the extended nature of the [N {\small II}] emitting 
region (see van Loon et al., in prep.).  

The distance and luminosity of AFGL 4106 are uncertain,
which makes it hard to decide directly whether the progenitor 
was an AGB star or a RSG. The luminosity of 
$\mbox{L}_\star/\mbox{L}_\odot \approx 10^{4} 
\times d^2$ (with $d$ in kpc), suggests that it is a luminous object, 
which until 2~kpc is a post-AGB star, however if this object is located 
at a larger distance a post-RSG nature cannot be excluded.
We will present evidence that AFGL~4106 is likely a post-RSG.
The optical to sub-millimetre energy distribution of AFGL~4106 was analysed in
detail by HKV and G\"urtler et al. (1996, hereafter GKH). Both
studies find a very high mass loss which recently stopped, however with
rather different density distributions in the shell.

AFGL~4106 was selected for observations with the Infrared Space
Observatory (ISO) in the guaranteed time of the Short Wavelength 
Spectrometer (SWS) consortium because of its classification as a
transition object, and because of its very high infrared flux levels.
The high-quality ISO data allow a detailed determination of the dust
composition and its distribution in the circumstellar shell, which 
may yield new insight in the origin and evolution of this remarkable object.
Waters et al. (1996) studied the 30-45 $\mu$m SWS spectrum of AFGL~4106
and discovered
weak narrow emission bands that can be attributed to crystalline
silicates. Here we present a full inventory of the solid state emission bands 
and their identification. For the first time a full radiative transfer
model of a CSE including crystalline silicates is presented.

This paper is organised as follows; In Sect.~2 we report on the 
observations and we discuss the data reduction procedures.
In Sect.~3 we present the stellar near infrared spectrum (2.4 -- 5 $\mu$m),
which shows evidence for the binary nature of this object.
In Sect.~4 the infrared excess at 5 -- 195 $\mu$m, caused by the
dust shell, is analyzed. We identify the dust composition 
using the different solid state features detected in this part of the spectrum.
In Sect.~5 we present a radiative transfer fit to the spectrum.
In Sect.~6 we discuss the results of this paper and in 
Sect.~7 we summarize our main conclusions.

\section{The observations and data reduction}

\subsection{ISO observations}

Spectra of AFGL~4106 were obtained using the Infrared Space Observatory (ISO,
Kessler et al. 1996). Three spectra were taken with the 
Short Wavelength Spectrometer (SWS, de Graauw et al. 1996) and one 
with the Long Wavelength Spectrometer (LWS, Clegg et al. 1996, 
Swinyard et al. 1996).
An overview of the ISO observing log can be found in Table~\ref{obslog}.
Part of spectrum II has already been shown in Waters et al. (1996).
Because of its better signal to noise, we will use spectrum~IV for our 
discussion in this paper, but information from the other spectra has
been used in the reduction of spectrum~IV.

\begin{table}
\begin{flushleft}
\begin{tabular}{lllll}
\hline
Nr.	& Instrument +		& Date 		& orbit	& t$_{\rm obs}$ \\
	& Observing mode	&  		&	& (sec)   \\
\hline
I 	& SWS AOT01 speed 1 	&  16-01-96 	&60 (PV)& 1172 \\
II 	& SWS AOT01 speed 2 	&  29-02-96 	& 104  	& 1944 \\
III 	& LWS 01            	&  29-02-96 	& 104 	& 509  \\
IV 	& SWS AOT01 speed 3 	&  22-07-96 	& 249 	& 3486 \\
\hline
\end{tabular}
\caption{ISO observing log for AFGL~4106 (PV = Performance Verification phase)}
\label{obslog}
\end{flushleft}
\end{table}

\subsubsection{SWS}
\label{sec:sws}

The SWS spectra (2.36 -- 45.3 $\mu$m) consist of 12 sub-spectra 
with slightly overlapping wavelength coverage
and each observed by 12 detectors.
Each spectrum is scanned twice, the so called up- and down-scan 
(de Graauw et al. 1996).
The spectra were reduced using the SWS off-line analysis pipeline,
version 6.0. For a description of flux and wavelength calibration
procedures, we refer to Schaeidt et al. (1996) and Valentijn et al. (1996).
In the reduction process the calibration files available in 
April 1998 were used. Judging from the overlap regions of the different 
sub-bands, further improvement is still possible, but will not influence 
the overall conclusions of our results.

The main fringes in the 12.0 -- 29.5 $\mu$m part of the spectrum 
were removed using the InterActive (IA) data reduction package 
routine {\em fringes}. Major irregularities due to
glitches and large drops or jumps in each detector were removed by hand. 
A comparison between the different detectors scanning the same wavelength 
region was used to determine the location of such jumps.
Detectors with a poor performance at the time of observations
were removed.

Several procedures were performed on each sub-band separately to combine the 12
detectors. A sigma clipping procedure was used to remove 
deviating points. The flux level of the 12 detectors was
brought to the average level of the 12 detectors. Finally each sub-spectrum
was rebinned to a resolution of $\lambda / \Delta \lambda = 750$.

The individual sub-spectra, when combined into a single spectrum, can 
show jumps in flux levels at band edges due to imperfect flux calibration
or dark current subtraction.
Based on the observed jumps, downwards with increasing aperture size,
we assume that the flux differences are not due to the significant
difference in beam size. Indeed, our ground-based
10 $\mu$m image (see Sect.~\ref{imaging}) 
shows an angular size well within the SWS apertures. 
We have adjusted the different subbands, according to the expected
source of discrepancy, to form a continuous spectrum.
Note that all the shifts are well within the photometric
absolute calibration uncertainties (Schaeidt et al. 1996). 
The 29.5 -- 31 $\mu$m section is severely
affected by memory effects of the detectors.
Therefore we have removed this part from the final spectrum.

\begin{figure}[ht]
\centerline{\psfig{figure=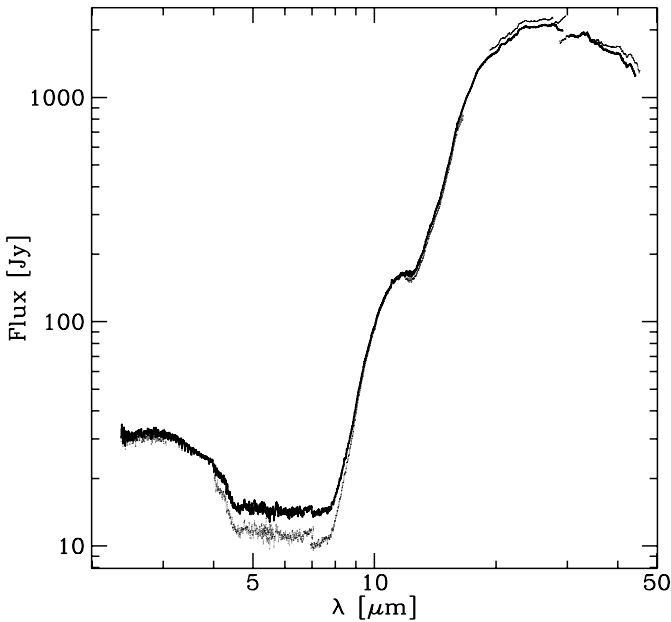,width=95mm}}
\caption[]{The final SWS spectrum (thick line), with the different sub-bands 
attached to each other, compared with the original outcome of the data reduction
(dots).}
\label{rawspec}
\end{figure}

In Fig.~\ref{rawspec} we show the final SWS spectrum and we compare
this to the original (without band matching) data of spectrum~IV.
A comparison of the final SWS spectrum with the original data of spectrum~II
gives a match of similar quality.

\subsubsection{LWS}

Spectrum~III is an LWS01 full spectral scan (43 -- 195 $\mu$m).
The spectrum was sampled with one point per spectral
resolution element using fast scanning. An off-set position spectrum
was also taken with the same instrumental parameters, in order to correct for
background contributions. The
spectrum was reduced using the LWS off-line analysis pipeline, version 6.0.  
See Table~\ref{obslog} for an overview of the
observing mode and date of observation. 

The Auto Analysis results of both source and background spectrum were 
then processed using the ISO Spectral Analysis Package (ISAP). First 
a median clipped average was calculated of all the spectral scans for 
one detector. These averaged scans were then fitted together using the 
scan on the LW1 detector as reference. The resulting spectrum was
rebinned to a standard wavelength scale with a resolution of 250 using
a flux conserving rebinning routine.  
Finally, the on-source and off-source spectrum were subtracted.
The final SWS + LWS spectrum is presented in Fig.~\ref{IRASLRS}.

\subsubsection{Comparison with IRAS data}
\label{sec:compare}

\begin{figure}[ht]
\centerline{\psfig{figure=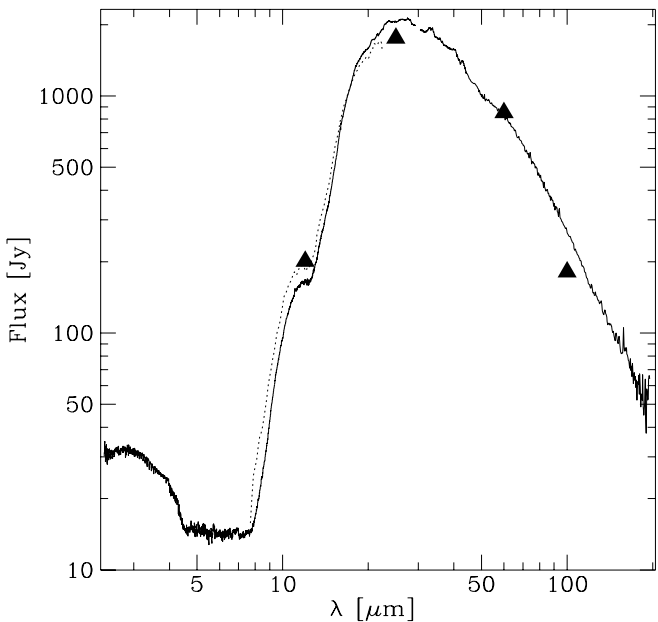,width=95mm}}
\caption[]{The final SWS and LWS spectrum of AFGL~4106 (solid line) 
compared with the IRAS-LRS (dotted line) spectra and the IRAS broad band 
fluxes (triangles)}
\label{IRASLRS}
\end{figure}

In Fig.~\ref{IRASLRS} we show the complete spectrum together with the IRAS
broad band fluxes.
We compared the resulting 2.4 -- 195 $\mu$m spectrum with IRAS data by 
convolving the ISO spectrum with the IRAS photometric band transmission curves
(IRAS explanatory supplement 1988),
the resulting fluxes are given in Table~\ref{IRAS}.

\begin{table}[ht]
\begin{flushleft}
\begin{tabular}{rcc}
\hline
Band	& IRAS-flux 	& Folded ISO-flux\\
($\mu$m)& (Jy) 		& (Jy)\\
\hline
12 	& 200.8 	& 169\\	
25 	& 1755 		& 1883\\	
60 	& 851.8 	& 910 \\	
100	& 181.1 	& 283\\	
\hline
\end{tabular}
\caption{The comparison with the IRAS broadband flux density}
\label{IRAS}
\end{flushleft}
\end{table}

The differences found for the 25 and 60 $\mu$m flux densities are well within 
the error that IRAS, SWS and LWS quote for their flux uncertainty. 
This also holds for the 12 $\mu$m flux density, however
one should realize that discrepancies in this 
wavelength regime might be related to the memory effects found in  
band~2C both for this source (as was found as a difference 
in the up- and down-scan) 
as well as in the calibration sources.

The 100 $\mu$m IRAS flux density (181.1 Jy) is considerably lower than 
the flux density in the LWS spectrum (283 Jy) and is outside the errors
quoted for IRAS and LWS. We have inspected the raw IRAS data covering
the position of AFGL~4106 for bad scans or steep background gradients,
which may cause the difference between both observations. We found no
bad scans but there is a conspicuous gradient in the galactic
background radiation. However, the IRAS 100 $\mu$m flux is corrected for
this background. Although the contribution of the background to the LWS
continuum flux was found to be minor,
the presence of the [C {\small II}] line at 158 $\mu$m, 
likely to be of interstellar origin, 
in the continuum subtracted spectrum suggests that the background has not
been subtracted correctly. At present, the difference between 
IRAS 100~$\mu$m and LWS is not understood.
In Fig.~\ref{IRASLRS}, we also show the comparison of the IRAS Low Resolution
Spectrograph (LRS) and the final spectrum. 
The agreement of the shape and flux levels of the LRS and the SWS
spectra is satisfactory.

\subsection{10 $\mu$m imaging at the ESO 3.6m telescope}
\label{imaging}

\begin{figure}[ht]
\centerline{\psfig{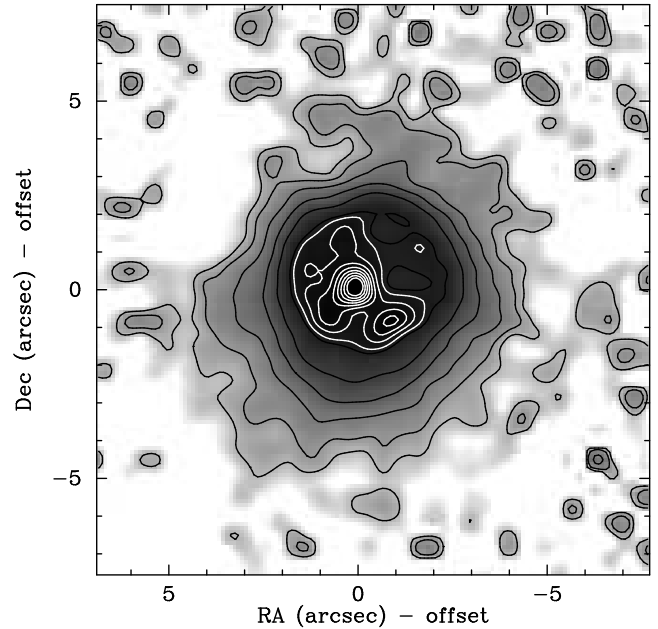}}
\caption[]{A logarithmic scale plot of the N-band image of AFGL~4106.
The black line contours start at $\sigma = 1$ and are steps of factors 3,
the white line start at $\sigma = 300$ and increase in steps of 100 up to 1000.
North is up and East is to the left.}
\label{timmi_in}
\end{figure}

On 23 February 1994 we used the mid-infrared camera TIMMI
(see K\"{a}ufl et al. (1994) for a description of the instrument) at the ESO
3.6~m telescope on La Silla, Chile, to image AFGL~4106. We used a broad
band N filter, centered at 10.1~$\mu$m. Standard infrared techniques were
used, i.e. chopping and nodding with a throw of  23.5$^{\prime\prime}$ ,
somewhat larger than the $21.55^{\prime\prime}$ field of view. We integrated
for 31 minutes on the source. We chose a small pixel scale of
$0.336^{\prime\prime}$ to avoid undersampling of the Point Spread Function 
(PSF). 
We observed the standard stars $\alpha$~Hya, $\epsilon$~Mus and $\alpha$~Car,
to allow for absolute flux calibration and to derive the 
PSF (FWHM $\approx 1.2''$) of the telescope plus instrument. 
We reached the
diffraction limit of $0.6^{\prime\prime}$ by applying a deconvolution
algorithm (Lucy 1974) within the MIDAS reduction software package. 
This deconvolution method has as drawback that it focuses the noise in the 
background, which is therefore quite spiked.
The deconvolved image is shown in Fig.~\ref{timmi_in}. 
The integrated instrumental\\
N-band magnitude (not corrected for color 
differences etc.) is $N=-1.19 \pm 0.05$. This corresponds to $\approx$ 120 Jy.
The convolution of the N-band sensitivity with the ISO spectrum
gives a flux of 100 Jy. The difference between both flux levels is probably
caused by detector memory effects in the ISO band 2C wavelength region, 
and by the slightly extended nature of the nebula.
The brightest, unresolved emission peak in the image is likely to be centered 
on the central object.
The nebula, especially the outer parts, seems to have an oval to 
box-like shape.
The inner part of the emission complex is characterized by
structure on (sub-)arcsecond scale in the form of local enhancements
of 10~$\mu$m emission. Whether this also reflects a clumpy density
structure is not certain.
The North-West part of the nebula is fainter compared to the rest of the 
nebula at 10~$\mu$m, and seems slightly more extended.
The whole nebula and especially the inner part has a North-West -- South-East 
symmetry axis.
The unresolved central peak contains roughly 13\% of the total flux,
which is about 15 Jy.
The total emission extends out to $\approx 5^{\prime\prime}$ from the center. 

An arc of H$\alpha$ emission, extending from 
roughly North-East to South-West clockwise, has been found 
(van Loon et al., in prep.), and is located just outside the edge of the 
extended dust emission in the N-band.  This arc coincides with the low 
surface brightness emission region in the N-band image.

\section{The 2.4 -- 5 $\mu$m part of the spectrum; A binary revealed}
\label{atmosphere}

The spectral energy distribution (SED) can naturally be divided into two 
parts. For wavelengths shorter than
5 $\mu$m the photosphere radiation dominates the spectrum. 
Above 5 $\mu$m the thermal radiation of the dust dominates the spectrum. 
This IR-excess will be discussed in Sect.~\ref{sec:dust}.
Here, we will present evidence that the central object is a {\em binary}
system, consisting of an M and A-F type star of almost equal luminosity.

\begin{figure}[ht]
\centerline{\psfig{figure=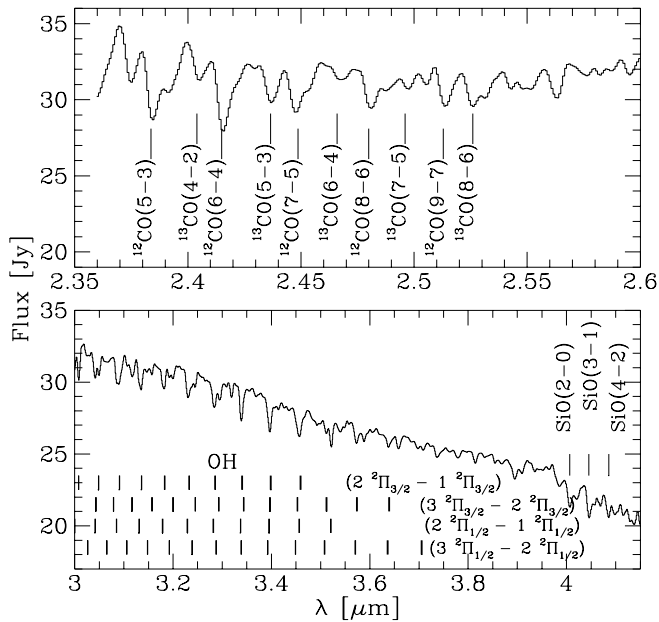,width=95mm}}
\caption[]{Ro-vibrational band heads of different molecular species, 
proving the existence of a cool companion to HD~302821}
\label{molband}
\end{figure}

The $2.4-5 \mu$m part of the spectrum is shown in Fig.~\ref{molband}.
It is dominated by
strong absorption from several band heads of ro-vibrational molecular bands, 
such as the CO first overtone and fundamental, the SiO first overtone, 
H$_2$O ($\nu_1$ and $\nu_3$) and OH fundamental.
The strength and excitation levels of these bands are not compatible with
those expected for a G-type supergiant spectrum, but suggest a cooler star.
From the shape and the level of the first overtone 
ro-vibrational $^{12}$CO bands we derived a temperature 
of $\approx 4000$ K (see e.g. $\alpha$-Boo in Decin et al. 1997).
This high temperature together with the required column density cannot be 
caused by circumstellar gas in a detached
(dusty) envelope. 
Also the presence of the high excitation lines of
gas-phase SiO, which cannot be in a dusty envelope,
suggests that we are dealing with a cool star.
Meanwhile van Winckel et al. (in prep.) took a high resolution spectra of 
AFGL~4106 in the optical and around 1$\mu$m and found a warm star of 
T $\approx 7000 - 7500$~K, dominating in the optical and clear evidence of both
the warm and the cool star around 1 $\mu$m, where both stars appear to be 
equally luminous.

\begin{figure}[ht]
\centerline{\psfig{figure=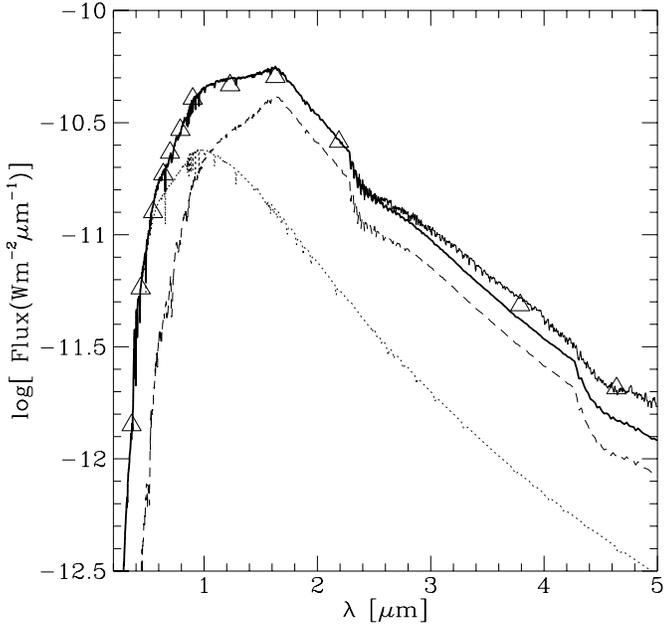,width=95mm}}
\caption[]{The reddened ($E(B-V) =1.2)$ 
composite spectrum (thick line) of the hot star (dotted line) 
and the cool star (dashed line) compared with the broadband photometry 
(triangles) and the ISO-data (thin line). The error in the photometry
is much smaller than the size of the triangles.}
\label{fig:binary}
\end{figure}

In order to verify the binary hypothesis, we
have combined the optical and near-IR photometry of AFGL~4106
(HKV, Garc\'{\i}a-Lario et al. 1994, 1997) together with our
ISO data.
We used Kurucz (1991) model atmospheres to fit the broad-band photometry
up to 2.2 $\mu$m. This ensures that the fit is not
contaminated by dust emission. For our fitting procedure we assumed that 
the mean interstellar extinction law can be used to describe both the 
circumstellar and the interstellar dust extinction. We will come 
back to this point in Sect.~\ref{outcome}.
We were not able to fit the data with a single star, but found a good fit
if we assumed a binary.
A best fit was found with temperatures of $7250\pm 250$K and $3750\pm250$K
for the hot and cool component respectively, while the luminosity ratio 
(L$_{\rm warm}$/L$_{\rm cool}$) = 1.8. For both components we independently 
found the same reddening of $E(B-V) =1.2 \pm 0.1$ mag.
The fit together with the observations is shown in Fig.~\ref{fig:binary}.

The relatively small luminosity difference between both stars suggests
that {\em both} components are evolved, i.e. the M star is a giant
or supergiant. A more detailed comparison of the ISO 2.4-5 $\mu$m
spectrum with that of other M giants and supergiants indicates
a supergiant nature of the spectrum (Van Winckel et al., in prep).

The flux from the central part of the ground-based 10 $\mu$m image is 
about 15 Jy.
This is much larger than the flux expected from the stellar 
atmospheres of both stars in the system (3 Jy).
This suggests that there is additional circumstellar material
very close to the binary, probably hot dust.
It is not clear whether this material was ejected at the same time
as the dust and gas in the detached shell, or whether this represents a later 
phase of mass loss. The additional circumstellar material could also originate
from the M-star in the system.
Note that the flux of the second brightest blob, SW of the central peak, is
still about 7 Jy, which is also enough to contain one or even both stars.
However one should keep in mind that both stars show the same reddening, 
suggesting that they are located behind the same dusty material
and that more clumps (e.g. SE, E and NNE of the central peak) are present.

\subsection{Interstellar and circumstellar extinction}
\label{extinction}

The large reddening of AFGL~4106 is illustrated by the 
prominent diffuse interstellar bands (DIB's) in the optical spectrum
of AFGL~4106.
\begin{figure}[ht]
\centerline{\psfig{figure=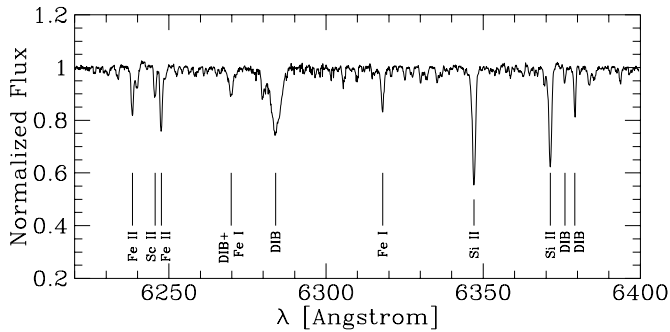,width=95mm}}
\caption[]{High-resolution optical spectrum of
AFGL~4106. The prominent interstellar DIB's are indicated together with the 
main photospheric atomic absorption lines. The spectrum has been corrected for 
the presence of telluric lines.}
\label{dibs}
\end{figure}
In Fig.~\ref{dibs} we display part of the optical spectrum taken by
van Winckel et al. (in prep.) corrected
for telluric lines. The broad band at 6284~$\rm \AA$, which is a blend of
different components, is strong,
but also the narrow DIB's at 6376 and 6379~$\rm \AA$\, can easily be
recognized. Since the carriers of the DIB's are associated with the
carbon-rich component of the interstellar medium (ISM) and the
circumstellar material around AFGL~4106 is oxygen-rich, the DIB's are likely
to be purely from interstellar origin and can be used to estimate the
interstellar component of the total reddening of AFGL~4106. 
The strength of the DIB's scales only roughly with the
reddening ($E(B-V)$) (see Herbig 1995), 
therefore our quantisation of the interstellar reddening is only indicative.

We compared the strength of the DIB's in AFGL~4106 with the DIB strength
in BD+63$^{\circ}$1964 ($E(B-V) = 1.01$~mag., Spectral type B0II) 
(Ehrenfreund et al. 1997) 
and derived a second estimate of the reddening by using the 
equivalent width to reddening relation derived by Jenniskens \& D\'esert
(1994) from a detailed
analysis of 4 reddened early-type stars. We limited ourselves to
narrow DIB's without photospheric contamination since the broad
DIB's in our spectrum are composite and the equivalent width
determination most inaccurate. The equivalent
width measurements are given in Table~\ref{blabla}. 
\begin{table} 
\begin{tabular}{llll}
\hline
$\lambda$ & AFGL~4106 & BD+36$^{o}$1964 & W$_{\lambda}$/E(B-V) \\
(\AA)	& EW(\AA)	& EW(\AA)	& \\
& & & \\ \hline
5849.78 & 72. &  121.& 48. \\
6196.19 & 76. &  78. & 61. \\
6325.10 & 33. &      & 18. \\
6330.42 & 42. &  16. & 18. \\
6376.02 & 37. &  28. & 26. \\
6379.27 & 140.&  176.& 78. \\
6425.72 & 22. &  25. & 19. \\
6613.72 & 319.&  316.& 231. \\
6660.64 & 51. &  57. & 51.  \\
6993.18 & 80. &      &116.  \\
7224.18 & 252.&      &259.  \\
\hline
\end{tabular}
\caption{Equivalent width of a selection of DIB's in
the optical spectrum of AFGL~4106 compared to the strength of same DIB's in
BD+63$^{o}$1964 (Ehrenfreund et al. 1997) and the DIB's-reddening
relation of Jenniskens \& D\'esert}
\label{blabla}
\end{table}             
One can see that the DIB's in AFGL~4106 and BD+63$^\circ$1964 are very similar.
We estimate the reddening using a linear 
least mean square analysis and found $E(B-V) = 1.07$ mag. 
by using the mean relation of Jenninskens \& D\'esert and a similar
value $E(B-V) = 0.94$ mag.
by using BD+63$^\circ$1964 as a standard.
We conclude that the interstellar 
reddening towards AFGL~4106 is $E(B-V) = 1.0 \pm 0.2$ mag.
This is in very good agreement with the value 
derived by van Loon et al. (in prep).

The case of AFGL~4106 offers a unique opportunity to measure the {\em
internal extinction} within the cicrumstellar dust shell. This is because
the star and the dust shell are surrounded by a shell of ionized gas,
detected in H$\alpha$ and the [N~{\small II}]~line. These emission lines were
reported by Garc\'{\i}a Lario et al. (1994), and subsequent imaging of the
spatial extent of the emission lines by Van Loon et al. (in prep.)
showed that the lines originate from a roughly spherical shell of gas on
the periphery of the dust shell. The [N~{\small II}]~lines were found to
be split, with the red-shifted component systematically weaker than the
blue peak. Van Loon et al. (in prep.) interpret this difference in
intensity as a result of internal extinction in the dust shell, and 
assuming a reddening law which is the same for the interstellar as well as
for the circumstellar dust, they found a circumstellar reddening towards the 
central star of 
$E(B-V) = 0.22 \pm 0.05$~mag. We conclude that sum of the DIB (interstellar)
extinction of about $E(B-V) = 1$~mag. 
and the [N~{\small II}] (circumstellar) extinction
of about $E(B-V) = 0.2$~mag. agrees well with the derived total line-of-sight
extinction of $E(B-V) = 1.2$~mag. based on the SED.

From the total spectrum one can derive the total flux, 
infrared excess and therefore another estimate of the interstellar
and circumstellar absorption towards AFGL~4106.
We define the infrared excess as the integrated flux under
the energy distribution corrected for the stellar contribution.
The distance to AFGL~4106 is not known, so we can only
determine a distance dependent luminosity, which is roughly 
$10.8 \times 10^3 \, d^2 $L$_\odot$, with $d$ the distance in kpc.
The IR-excess contributes about 75\% 
($8.0 \times 10^3 \, d^2 $L$_\odot$) to the total luminosity.
Which leaves $2.8 \times 10^3 \, d^2 $L$_\odot$ for the luminosity
directly from the reddened photospheres of both stars.
Assuming an $E(B-V)$ of 1.2 mag. for both stars, 
taken from the fit to the broad-band
photometry, results in a dereddened luminosity of 
$\approx 7.4 \times 10^3 \, d^2 $L$_\odot$ for the hot component
and $\approx 4 \times 10^3 \, d^2 $L$_\odot$ for the cool one.
This implies a total ``loss'' of stellar energy by absorption and scattering of
$\approx 8.6 \times 10^3 \, d^2 $L$_\odot$/s. 
This would imply that, assuming a spherical
dust shell and assuming that all absorbed energy is 
re-radiated in the infrared, more than 90\% of the extinction towards 
AFGL~4106 is circumstellar, and thus the
contribution of the interstellar extinction would be negligible.
This is not in agreement with the estimate of the interstellar extinction
derived from the depth of the DIBs (see above). 
In Sect.~\ref{outcome} we will demonstrate that this discrepancy
is resolved by the non standard extinction properties of the CSE.

\section{The 5 -- 195 $\mu$m part of the spectrum; The dust shell}
\label{sec:dust}

\begin{figure}[ht]
\centerline{\psfig{figure=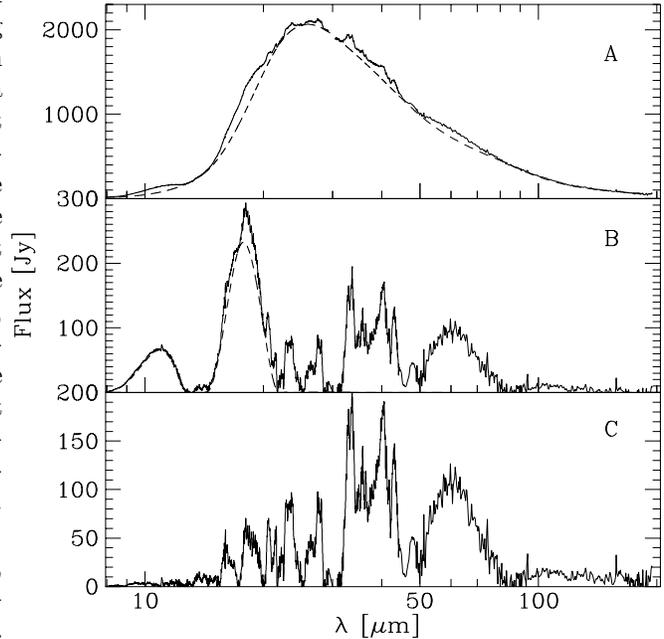,width=95mm}}
\caption[]{The SWS and LWS spectrum of AFGL~4106 with the spline
fit continuum. The result of the subtraction of the
spline fit continuum (dashed line in A)
is shown in graph B. The result of the removal of the broad silicate features
(dashed line in B) is given in graph C.}
\label{contsub}
\end{figure}

At wavelengths longwards of about 5 $\mu$m thermal emission from 
the circumstellar dust starts to dominate the spectrum
(see Fig.~\ref{IRASLRS}). Between 5 and 7.8~$\mu$m the spectrum is 
flat at a level of approximately 15~Jy. 
Around 7.8~$\mu$m a steep rise is found with a flattening around 12~$\mu$m.
This prominent structure is caused by an increasing continuum together
with the 10 $\mu$m amorphous silicate feature (Si-O\\
stretching mode). 
At 13~$\mu$m the flux increases again, with at around 18~$\mu$m  the
second peak of the amorphous silicate feature (O-Si-O bending mode).
The peak of the infrared excess is found around 25~$\mu$m.
Superimposed on this continuum with the broad features are narrow features,
especially in the SWS part of the spectrum. In most cases they can be 
attributed to crystalline silicates.

In contrast with the SWS spectrum with its sharp spectral features, the LWS 
spectrum of AFGL~4106 is very smooth. Emission lines are 
present in the on-source spectrum at 88.356 ([O {\small III}] line) and
157.741 $\mu$m ([C {\small II}] line). However, since they are also present in 
the background spectrum, they cancel out almost completely in the final 
spectrum. Although the warm component in the binary is not hot enough to 
produce the [O {\small III}]-line, the presence of 
the [O {\small III}]-line 
may not be surprising given the line of sight (Carina) and the projected
vicinity of several very luminous hot stars,
which emit enough ionizing radiation to produce the [O {\small III}]-line. 
We find no evidence for emission lines in the SWS and LWS infrared spectra
from the ionized nebula surrounding AFGL~4106 (detected at optical 
wavelengths), which should be within the ISO beam. 
Clearly the dust continuum swamps any line emission from that region. 

\subsection{The continuum subtraction of the SWS and LWS spectra}
\label{sec:contsub}

\begin{figure*}[ht]
\centerline{\psfig{figure=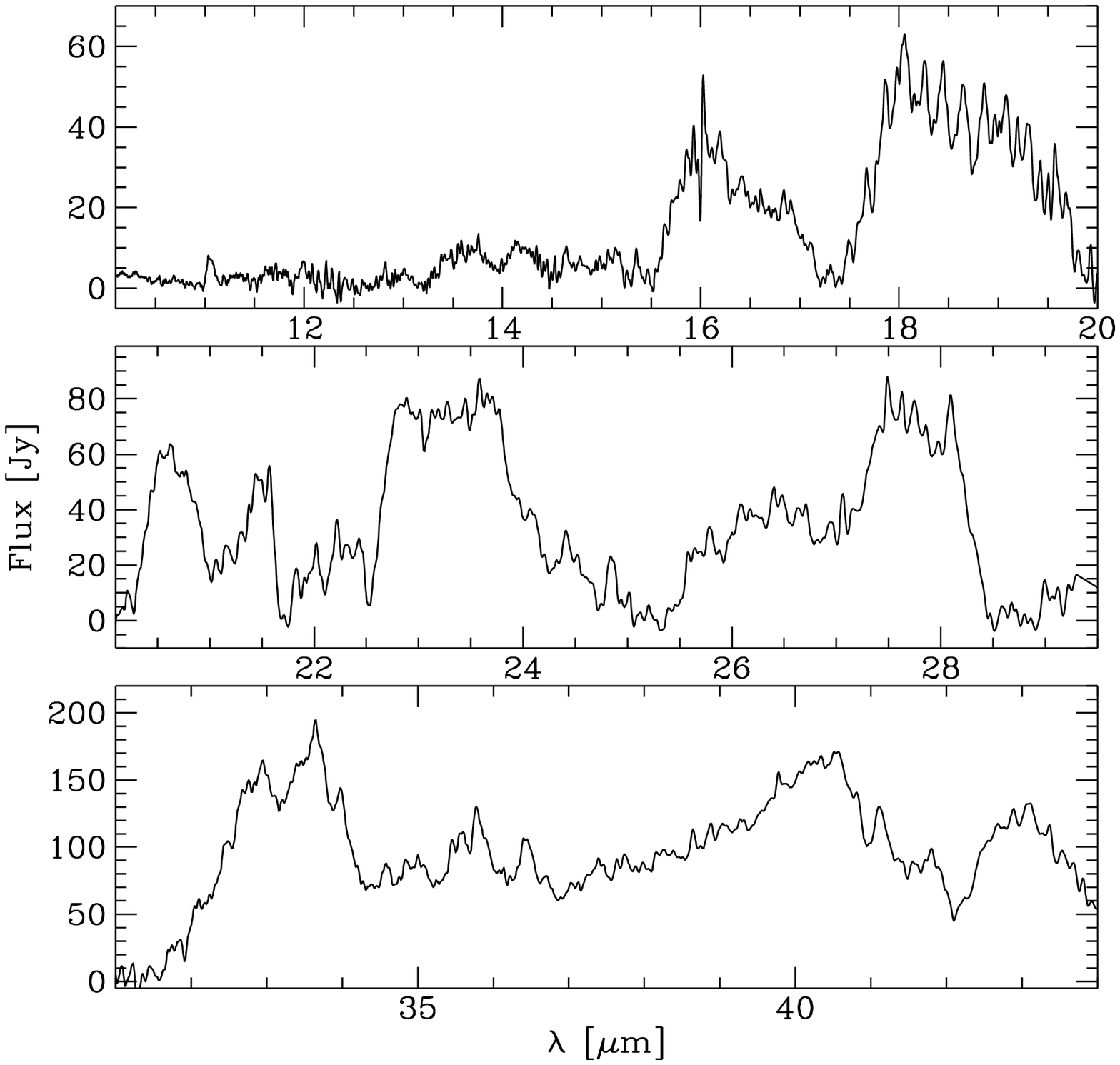,width=180mm}}
\caption[]{The 10 to 44 $\mu$m region of the continuum subtracted spectrum
enlarged. Note the richness of solid state features.}
\label{zoomin}
\end{figure*}

In order to show the presence of the spectral features more clearly, 
we have subtracted a continuum by fitting a spline curve to the 
spectrum. Note that the spline fit continuum has no physical meaning and 
is only used to enhance the sharp solid state features on top of the spectrum.
The continuum subtracted spectrum
is shown in Fig.~\ref{contsub} panel~B.
The amorphous silicate features dominate the continuum subtracted 
spectrum below 20 $\mu$m. In order to show the narrow bands
more clearly, we have created a second continuum subtracted spectrum,
where we have excluded the broad amorphous silicate bands
(see the dashed line in Fig.~\ref{contsub} panel B and the result in 
panel C).
We find a wealth of solid state structures, which we show in more detail
in Fig.~\ref{zoomin}.
Some residual fringing is still visible at 18-20 $\mu$m.

\begin{table*}
{\small
\begin{tabular}{lccllll}
\hline
\multicolumn{1}{l}{$\lambda$} & 
\multicolumn{1}{c}{F$_{\rm band} $}	& 
\multicolumn{1}{c}{FWHM}	& 
\multicolumn{1}{c}{FWHM/$\lambda$} & 
\multicolumn{1}{c}{I$_{\rm peak}$/I$_{\rm cont}$} &
\multicolumn{1}{l}{Identification} & 
\multicolumn{1}{l}{Comments} \\
\multicolumn{1}{c}{$ \mu$m}	   & 
\multicolumn{1}{c}{$10^{-14}$ Wm$^{-2}$}  &
\multicolumn{1}{c}{$\mu$m} &	&	&	\\
\hline
10.8 &$4\times10^2$& 2.5 &  0.23 & 1.8   & amorph. silicate      &    \\
16.1    & 40.   & .7    & 0.043 & 1.06  & cryst. forsterite   	& on the slope of the 17.6 $\mu$m feature \\
        &       &       &       &       &       		& and measured together with the 16.8 $\mu$m peak \\
16.8    & 15.   & .6    & 0.036 & 1.03  & unidentified          & on the slope of the 17.6 $\mu$m feature \\
        &       &       &       &       &                       & and measured together with the 16.1 $\mu$m peak \\
17.6 &$5\times10^2$& 3.  & 0.17  & 1.2   & amorph. silicate      &     \\
18.1    & 75.   & 1.0   & 0.055 & 1.07  & cryst. enstatite   	& on the slope of the 17.6 $\mu$m feature \\
        &       &       &       &       &                       & and measured together with the 19.2 $\mu$m peak \\
19.2    & 50.   & 1.0   &  0.052& 1.045 & cryst. enstatite and 	& on the slope of the 17.6 $\mu$m feature \\
        &       &       &       &       & cryst. forsterite 	& and measured together with the 18.1 $\mu$m peak \\
20.6 	&  20.  & 0.48  & 0.023 & 1.035 & cryst. enstatite ?	& \\
21.5 	&   7.3 & 0.25  & 0.012 & 1.024 & cryst. enstatite and	& Sharp edge caused by responsivity\\
      	&       &       &       &       & cryst. forsterite? 	& \\
22.9 	& 25.   & 0.5   & 0.021 & 1.04  & unidentified          & measured together with the 23.6 $\mu$m peak \\
     	&       &       &       &       &                       & Sharp edge caused by responsivity\\
23.6 	& 30.   & 0.7   & 0.030 & 1.04  & cryst. forsterite  	& measured together with the 22.9 $\mu$m peak \\
24.1  	&  1.8  & 0.16  & 0.0066& 1.009 & unidentified          & On the slope of the 23.6 micron feature \\
24.5    &  4.1  & 0.33  & 0.013 & 1.011 & cryst. enstatite?     & On the slope of the 23.6 micron feature \\
26.2  	& 16.   & 1.05  & 0.040 & 1.02  & cryst. forsterite  	&  \\
27.8 	& 44.   & 1.15  & 0.041 & 1.05  & cryst. forsterite and &  \\
      	&       &       &       &       & cryst. enstatite  	&  \\
32.8 	& 20.   &  .6   & 0.018 & 1.07  & cryst. enstatite  	& measured together with the 33.6 $\mu$m peak \\
      	&       &       &       &       &                   	& see remarks in text\\
33.6  	& 40.   & 1.0   & 0.030 & 1.08  & cryst. forsterite	& measured together with the 32.8 $\mu$m peak \\
34.0  	&   .96 &  .12  & 0.0035& 1.016 & cryst. enstatite ? 	& on the slope of the 33.6 $\mu$m peak   \\
35.8 	&  5.5  &  .54  & 0.015 & 1.03  & cryst. enstatite   	&  \\
36.5 	&  2.4  &  .28  & 0.0077& 1.02  & cryst. forsterite  	&  \\
40.4  	& 13.   & 1.1   & 0.027 & 1.05  & cryst. enstatite	&  \\
41.1 	&  1.1  &  .17  & 0.0041& 1.02 	& unidentified	  	& on the slope of the 40.4  $\mu$m peak  \\
43.1 	& 17.   & 1.2   & 0.028 & 1.07  & cryst. enstatite and	&  \\
      	&       &       &       &       & cryst. H$_2$O     	&  \\
47.8  	&  9.   & 1.9   & 0.040 & 1.036 & unidentified	 	&  \\
61  	& 68.   & 20.   & 0.33  & 1.05  & cryst. H$_2$O     	&  \\
\hline
\multicolumn{7}{c}{Spurious features}\\
\hline
11.07 	& 1.98  & 0.082 & 0.0074& 1.05  & & instrumental artifact \\
13.6    & 10.   & .5	& 0.037 & 1.04  & & measured together with the 14.2 $\mu$m peak \\
        &       &       &       &       & & instrumental artifact? \\
14.2    & 6.5	& .4	& 0.028 & 1.05	& & measured together with the 13.6 $\mu$m peak \\
        &       &       &       &       & & instrumental artifact? \\
\hline
\end{tabular}}
\caption[]{The characteristics of the dust features in the spectrum of 
AFGL~4106}
\label{tab:feat}
\end{table*}

Since the solid state bands have a wide range in width and strength, 
and often blend, we decided to measure the band strength with respect 
to a {\em local continuum}. We used a second order polynomial
fit to define a local
continuum in the final spectrum (i.e. not the continuum subtracted
spectrum shown in Fig.~\ref{contsub}~and~\ref{zoomin}), 
and fitted Gaussians to the
emission bands with respect to this local continuum. The very broad
amorphous silicate bands could not be fitted with a single Gaussian, and
we give the FWHM directly measured from the spectrum.
The results of the Gauss-fits can be found in Table~\ref{tab:feat}.
Due to our continuum subtraction method, it is likely that 
very broad features have been included in the continuum 
and therefore are not visible in the continuum subtracted spectrum.

In the following sub-sections, we briefly describe the continuum
subtracted spectrum.

\subsubsection{The 10-20 $\mu$m region}

The Si-O stretching and O-Si-O bending mode vibrations of
the amorphous silicates dominate the 10 - 20 $\mu$m
part of the spectrum. We note that the peak wavelength of the Si-O stretch
is significantly red-shifted compared to the bump found in most O-rich
(post-)AGB stars (Monnier et al. 1998) and the ISM. 
However, Little-Marenin and Little (1988, 1990)
find in LRS spectra an emission feature which peaks at a 
similar position in most S-, several MS- and some (5\%) M-type stars.
On top of the amorphous silicates are narrow emission features.
Careful inspection of the raw data suggests that there
are three instrumental artifacts at 11.07 and probably 13.6 and 14.2 $\mu$m,
but the other features at 16.1, 16.8, 18.1 and 19.2 $\mu$m
are of circumstellar dust origin. The
identification of most of these narrow peaks can be found in 
Table~\ref{tab:feat} and Sect.~\ref{sec:ident}.

\subsubsection{The 20-29.5 $\mu$m region}

The spectral region between 20 and 29.5 $\mu$m is shown in 
Fig.~\ref{zoomin} and is characterized 
by a wealth of emission features with a range of FWHM; we find 
narrow features (FWHM/$\lambda < 0.03$)  at 20.6, 21.5, 22.9 and 24.1
$\mu$m, and broader features (FWHM/$\lambda > 0.03$) at
23.6, 26.2, and 27.8 $\mu$m.
The sharp descent at the long wavelength side of the 21.5 $\mu$m feature
and rise at the short wavelength side of the 22.9 $\mu$m feature
are probably caused by an imperfect responsivity calibration. 

\subsubsection{The 29.5-45 $\mu$m region}

As stated in Sect.~\ref{sec:sws} we excluded 29.5 to 31 $\mu$m from the 
final spectrum, but we can say that we did not detect
narrow emission bands in this wavelength range.
The spectral region from 29.5 to 45 $\mu$m 
is dominated by a broad structure from about 
31 to 44 $\mu$m, on top of which 
emission features are present, see also Fig.~\ref{zoomin}. 
We find broad features at
33.6, 40.4 and 43.1 $\mu$m and narrow features at 32.8, 34.0, 35.8,
36.5 and 41.1 $\mu$m.
Waters et al. (1996) find a weak plateau from 32 to 37 $\mu$m
in the spectrum of several oxygen rich dust shells.
This plateau is also visible in AFGL~4106, although it is weak
and blends with the broad unidentified structure from 31 to 44 $\mu$m.   
There is also evidence for sub-structure in the 32-33 $\mu$m region, 
suggesting another weak band that contributes. 
This weak band at 32.0 $\mu$m is also seen in 
the SWS spectrum of NGC~6302 (Lim et al., in prep.) and in that of
HD44179 (the ``Red Rectangle'', Waters et al. 1998). 

\subsubsection{The 45-195 $\mu$m region, LWS}

At longer wavelengths, the spectrum is smooth. We find a narrow band at 
47.8~$\mu$m, and a broad emission centered at 61 $\mu$m (probably due to 
H$_2$O-ice). Also, the [C {\small II}]-line\\
at 157 $\mu$m is present;
this is likely due to imperfect cancellation of the on-source and off-source 
caused by the gradient found in
the background (see also the remarks in Sect.~\ref{sec:compare}).
For the moment, we ignore the [C {\small II}]-line emission.

\subsection{Identifications}
\label{sec:ident}

\subsubsection{Amorphous silicates}
\label{sec:amorph}

\begin{figure*}[ht]
\centerline{\psfig{figure=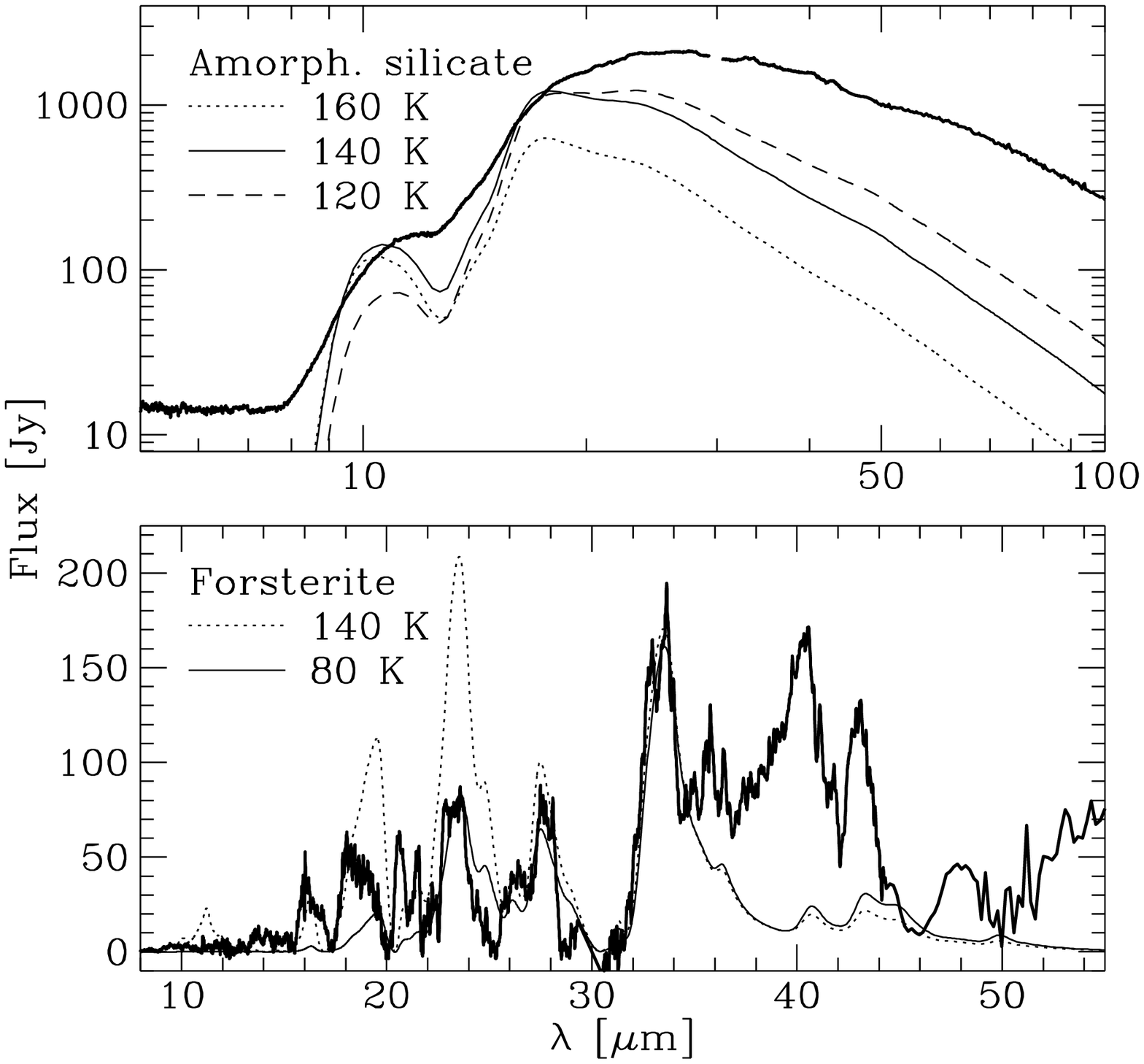,width=180mm}}
\caption[]{The emission spectrum of a 120, 140 and 160~K grain of 
amorphous olivine (Mg$_{0.8}$Fe$_{1.2}$SiO$_4$, Dorschner et al. 1995)
compared with the spectrum of AFGL~4106 (upper panel),
and the continuum subtracted emission spectrum of a 80 and 140~K 
crystalline forsterite (J\"ager et al. 1998) grain
compared with the continuum subtracted spectrum of AFGL~4106 (lower pannel).}
\label{dusttemp}
\end{figure*}

The broad emission features around 10 and 18 $\mu$m are attributed to amorphous
silicates. 
We have derived a rough estimate of the temperature of 140~K for the 
amorphous dust component using the amorphous olivine optical constants 
published by Dorschner et al. (1995)
(see also Fig.~\ref{dusttemp}). 
Such a temperature
estimate should be used with caution, since the features are on top of a steep
continuum and contribution from other species (e.g. FeO) can
affect the 10 over 18 $\mu$m band strength ratio. 
A similar temperature was found by fitting the shape of the dust spectrum 
with a dust emissivity Q$(\lambda) \propto \lambda^{-1}$.
This simple procedure even fits the long wavelength part very well but
is significantly too low from 15 to 50 $\mu$m.

\subsubsection{Crystalline olivines and pyroxenes}
\label{sec:cryst}

We use the laboratory transmission spectra of crystalline silicates 
as measured by J\"ager et al. (1998) and Koike \& Shibai (1998) to identify 
the rich solid state spectrum. The identifications are based on 
the match between the laboratory peak positions and observed peak positions
for the minerals of the olivine (Mg$_{2x}$Fe$_{(2-2x)}$SiO$_4$) and
pyroxene (Mg$_{x}$Fe$_{(1-x)}$SiO$_3$) family.
We find a very good match between the most (but not all)
broad emission bands and the laboratory data
for the pure Mg end member of the crystalline olivine solution serie, 
i.e. forsterite (Mg$_2$SiO$_4$).
Similarly, we can identify most narrow emission bands 
with the pure Mg end member of the crystalline pyroxene solution serie, i.e.
enstatite (MgSiO$_3$). Enstatite comes in two crystal structures, ortho- and clino-enstatite. However, apart from the 32.8 $\mu$m peak which is only found 
in ortho-enstatite (Koike \& Shibai 1998) and blends with the 33.6 $\mu$m
forsterite peak, the main strong peaks of 
clino- and ortho-enstatite are located at about the same position.  
For this reason we were not able to distinguish between these two forms in 
Table~\ref{tab:feat}.

\begin{figure*}[ht]
\centerline{\psfig{figure=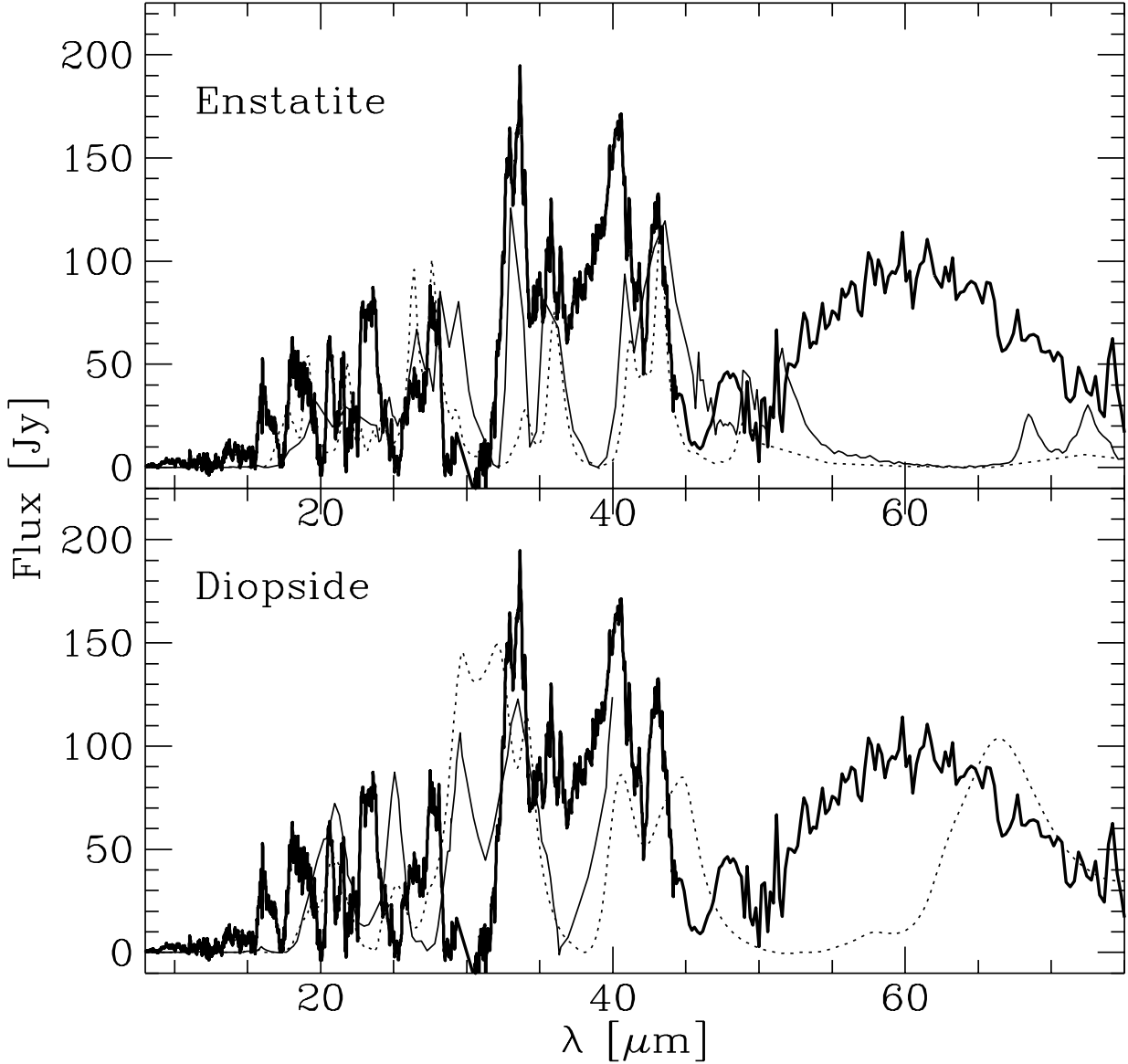,width=180mm}}
\caption[]{Upper panel: The continuum subtracted emission spectrum of a 
80~K crystalline ortho-enstatite grain from Koike \& Shibai (1998) 
(thin solid line) and from J\"ager et al. 1998 (dashed line) compared
with the continuum subtracted spectrum of AFGL~4106 (thick solid
line). Lower panel: The continuum subtracted emission spectrum of a 
80~K crystalline diopside grain from Koike \& Shibai (1998) 
(thin solid line) and a diopside measured by J\"ager (private communication)
compared with the continuum subtracted spectrum of 
AFGL~4106 (thick solid line).}
\label{ensttemp}
\end{figure*}

Apart from the wavelength of the bands, it is also important to
consider the relative strength of the emission bands in the identification
process, since it gives a rough estimate of the temperature.
To estimate the temperature we compared the continuum subtracted spectrum 
of AFGL~4106, with the continuum subtracted spectrum of forsterite
(and enstatite and diopside in Fig.~\ref{ensttemp}). The continuum subtraction
has been performed in the same way as it was done
for AFGL~4106 (i.e. a smooth spline fit curve without obvious kinks)
after multiplying the emissivity of the different dust
species with blackbodies of the given temperatures.
Since the peak of the features are at same wavelength for the two
temperatures, which had quite different continua, 
it shows that this continuum subtraction does not shift the peaks.
For forsterite we show in Fig.~\ref{dusttemp} the best fit to the 
continuum subtracted spectrum of AFGL~4106, assuming that the forsterite 
grains have a single temperature. The best fit temperature is about 80~K.
We stress that these temperature estimates are only approximate;
a full treatment of the radiative transfer is discussed in 
Sect.~\ref{modelling}.
The temperature of the amorphous dust is significantly higher than that of the 
crystalline component: a temperature of 140~K for the forsterite would have 
resulted in significant emission in the 11.2 $\mu$m band, which is not
observed. We conclude that amorphous and crystalline dust
are not thermally coupled and that they either have quite different optical 
properties, or are spatially distinct.

Unfortunately, the situation is rather disappointing for the pyroxenes.
The observed band strengths of AFGL~4106 poorly match the band strengths 
seen in the laboratory data published by J\"ager et al. (1998).
Therefore we cannot derive a reliable characteristic temperature for
the pyroxenes, based on these laboratory data.
We have also compared the band strengths of pyroxenes
measured by Koike \& Shibai (1998) to our AFGL~4106 data
which gives a slight improvement
of the match, especially in the 32.8 $\mu$m range (see Fig.~\ref{ensttemp}).
A temperature of 80~K gives again a reasonable fit to the data.
The difference between these two laboratory spectra might be caused by
slight differences in the chemical composition. 
The natural ortho-enstatite sample of 
J\"ager et al~(1998) is not the pure Mg end member of the crystalline pyroxene 
solution series but contains 2.1 mass \% FeO, 
while the data of Koike \& Shibai (1998) is 
of a synthetic sample of enstatite with some small impurities from LiO$_2$,
MoO$_3$ and V$_2$O$_5$.
Wheter this difference at 32.8 $\mu$m is due to the small amount of FeO,
which has influence on the peak over continuum ratio (see e.g. Fig.~2 and 
Fig.~3 of J\"ager et al. (1998)), or due to the metaloxides found
in the ortho-enstatite of Koike \& Shibai (1998) is not known yet, but we
will tentatively identify it with enstatite. 
For both the olivines and the pyroxenes the observed FWHM of the bands are 
significantly smaller than those measured in laboratory spectra.
It is interesting to note that both in the laboratory spectra and in our ISO 
data the olivine bands are broader than the pyroxene bands.

The main component for the plateau from 32 to 37 $\mu$m is unidentified yet. 
Its appearance and shape are influenced by the forsterite bands at 33.6 
and 36.5 $\mu$m and the enstatite band(s) at 35.8 (and 32.8) $\mu$m, 
which makes it
difficult to extract the unblended shape of this feature and therefore
its identification.

\subsubsection{Diopside}
\label{sec:diopside}

A third silicate that was checked for its presence is diopside (CaMgSiO$_3$). 
In Fig.~\ref{ensttemp} we show the comparison
of diopside with the spectrum of AFGL~4106. 
Although the two datasets (Koike \& Shibai, 1998; J\"ager, private 
communication) used for comparison show some differences, probably 
due to a somewhat different chemical composition, both datasets predict 
prominent emmision at 21 and 25 $\mu$m, which is not observed in our spectrum.
We conclude that the presence of diopside cannot be confirmed at present.

\subsubsection{Aluminum rich dust}
\label{sec:alum}

In Sect.~\ref{outcome} we will demonstrate that corundum 
($\alpha$-Al$_2$O$_3$)
is present in the dust shell of AFGL~4106. According to the condensation scheme
published by Tielens (1990) corundum can transform into melilite
(Ca$_2$Al$_2$SiO$_7$). It is not known by the authors if melilite will form 
below or above the glass temperature. 
Since we only have optical data for amorphous melilite (Mutschke et al. 1998)
we will concentrate on this.
Amorphous melilite shows broad features at 10, 15, 19 and 34 $\mu$m.
Because of the amorphous structure the features are very broad and
difficult to identify in the spectrum. However the
very broad (several $\mu$m) feature at 34~$\mu$m might be a good candidate for 
the broad structure from about 31 to 44 $\mu$m found in our
ISO-spectrum.

\subsubsection{Water ice}
\label{sec:water}

We have identified the feature at 61 $\mu$m, hereafter referred to as the 
60 $\mu$m feature, with crystalline H$_2$O ice.
Omont et al. (1990) identified a similar feature
observed with KAO in IRAS 09371+1212 (Frosty Leo) with crystalline 
H$_2$O ice, while
Barlow (1998) reports the presence of 60 $\mu$m emission in several dusty 
circumstellar shells. 
Laboratory data published by Bertie et al. (1969) show that 
crystalline H$_2$O ice has two strong bands, one at about 43 $\mu$m, 
the other at roughly 61 $\mu$m.
In the SWS spectrum we already found a broad and strong emission feature 
at 43.1 $\mu$m, which is probably a blend of crystalline enstatite and
crystalline H$_2$O-ice.
The fact that a 60 $\mu$m feature is present
strengthens the crystalline H$_2$O-ice identification.
\begin{figure}[ht]
\centerline{\psfig{figure=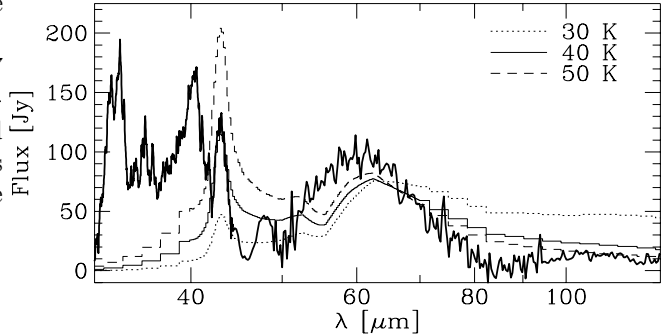,width=95mm}}
\caption[]{The emission spectrum of a 30, 40 and 50~K crystalline H$_2$O-ice
grain (Bertie et al. 1969) 
compared with the continuum subtracted spectrum of AFGL~4106.}
\label{icetemp}
\end{figure}
Assuming that the 43.1 $\mu$m feature is a blend of different features, 
including crystalline H$_2$O-ice, we have derived an upper limit of the 
temperature of only 40~K.
This is again much lower than the amorphous silicate temperature
but also significantly lower than the crystalline silicate temperature.
It cannot be excluded that the 60 $\mu$m feature is in fact a blend.
We note that the laboratory spectra of clino-enstatite (Koike \& Shibai 1998)
show a weak 60 $\mu$m bump. This would result in a higher temperature for the 
crystalline H$_2$O ice.
Still it points to another grain population which is not thermally coupled 
with the others.

\subsubsection{Iron oxides}
\label{sec:iron}

The identification of Mg-rich and Fe-poor crystalline silicates in
non-negligible abundances suggests that some Fe may be present
in the form of simple oxides. For spherical FeO grains 
the most important infrared feature longwards of 10 micron
is a relatively sharp 
feature at 20 $\mu$m, which is not present in our spectrum. However
FeO is very sensitive to shape effects (Henning \& Mutschke 1997). In the case
of a continuous distribution of ellipsoids (CDE, Bohren \& Huffman 1983) the 
20$\mu$m peak broadens 
significantly, the FWHM becomes roughly 8 $\mu$m,
and also shifts to 24 $\mu$m. 
This feature is temperature dependent (Henning \& Mutschke 1997) but only
in absolute strength and not in wavelength.
The broad feature is present in our total
spectrum, but hidden in the ``continuum'' in the continuum subtracted spectrum.
If FeO is present it is possible that 
Fe$_2$O$_3$ (hematite) and Fe$_3$O$_4$ (magnetite) are also present.
We have compared laboratory spectra of Fe$_2$O$_3$ (Steyer 1974) to our
observations. This oxide has strong bands in the 15-45 $\mu$m region,
that are sensitive to grain shape effects. We were unable to find
reasonable fits to the observed spectra for spherical and for a CDE mixture.
We conclude that Fe$_2$O$_3$ is not very abundant or absent in AFGL~4106.

Fe$_3$O$_4$ (magnetite) is quite featureless, except for a strong resonance 
at 17.5 $\mu$m and a weak one at 26.5 $\mu$m. These features are temperature 
dependent, not only in strength but also in shape and position and they have 
the tendency to shift to shorter wavelengths when the temperature decreases.
At 150~K the 26.5 $\mu$m feature is split in two features at 25.4 and 
26.7 $\mu$m, while the 17.5 $\mu$m feature shifts to 17.1 $\mu$m.
Since we do not see strong features in the wavelength range from
17.0 to 17.5 $\mu$m hardly any magnetite seems present.

\subsubsection{Unidentified features}
\label{sec:unident}

We find a narrow emission band at 47.8 $\mu$m; although the wavelength
of this feature is close to a spectral structure in the responsivity curve of 
LWS and a instrumental artifact can therefore not be excluded, quite convincing
evidence for its presence was given by Barlow et al. (1998) in the
spectrum of several oxygen-rich sources with crystalline silicates
at shorter wavelengths. 
According to the optical constants of Steyer (1974), 
almandine~(Fe$_3$Al$_{2}($SiO$_{4})_{3}$) peaks at
48 and 60 $\mu$m.
However, other peaks of almandine,
e.g. at 42 and 31.5 $\mu$m, are not observed. Therefore we
exclude this possibility.
We have not found a convincing identification for this 
band, but given its width it is likely caused by a crystalline silicate.

Several other peaks remain, which are not yet firmly identified 
(e.g. 16.8, 20.6, 22.9 and 24.1~$\mu$m).
It is therefore (very) likely that more dust components are present.

\section{Radiative transfer modelling}
\label{modelling}

In this section we model the observed SED of AFGL~4106 with a 
one-dimensional radiative transfer code (MODUST, Bouwman \& Waters, 1998;
Bouwman et al. in prep.; de Koter et al., in prep.).
From the structure seen
in the N-band images it is clear that this approach is not entirely correct.
Since AFGL~4106 is only marginally extended for ISO, we assume that 
diffraction losses can be neglected.

\subsection{Dust species}
\label{input}

We have tried to collect optical constants of the dust species, that are
identified in the previous section.
Unfortunately, in many
cases the optical constants are given only for a limited wavelength range,
and we had to combine data sets from different authors or extrapolate
to cover the full wavelength range required for our modelling.
The data sets used for the modelling can be found in 
Table~\ref{t:dust_input}.

We assume spherical grains and a power-law size distribution and we use Mie 
theory to calculate the optical properties of the grain population.
This approach allows us to calculate the emission for arbitrarily large grains.
As we will show below, large grains are abundant in the shell of AFGL~4106.

\begin{table*}[ht]
\begin{flushleft}
\begin{tabular}{llcl}
\hline
Dust species	& Chemical formula	&$\lambda$ ($\mu$m)	& reference \\
\hline
olivine		& Mg$_{0.8}$Fe$_{1.2}$SiO$_4$	& 0.1 -- 0.2	&extrapolated\\
(amorphous)	&			& 0.2 -- 500		& Dorschner et al. (1995)\\
		&			& 500 -- 1500		& extrapolated \\
\hline
forsterite	& Mg$_2$SiO$_4$		& 0.1 -- 5.0		& Scott \& Duley (1996)\\ 	
(crystalline)	&			& 5.0 -- 125		& Servoin \& Piriou (1973)\\
(all 3-axes)	&			& 125 -- 1500		&extrapolated\\
\hline
water ice	& H$_2$O		& 0.1 -- 1.25		&extrapolated\\
		&			& 1.25 -- 333		&Bertie et al. (1969)\\
		&			& 333 -- 1500		&extrapolated\\
\hline
iron-oxide	& FeO			& 0.1 -- 0.2		&extrapolated\\
		&			& 0.2 -- 500		&Henning et al. (1995)\\
		&			& 500 -- 1500		&extrapolated\\
\hline
corundum	& $\alpha$-Al$_2$O$_3$	& 0.1 -- 0.5		&extrapolated\\
		&			& 0.5 -- 400		&Koike et al. (1995)\\
		&			& 400 -- 1500		&extrapolated\\
\end{tabular}
\caption{The dust species used for the radiative transfer modelling}
\label{t:dust_input}
\end{flushleft}
\end{table*}

Since we identified both crystalline olivines as well as crystalline
pyroxenes in our spectrum, it is likely that also both components are 
present in the amorphous state. For the amorphous olivines
we applied the optical constants derived by Dorschner et al. (1995).
Unfortunately they only published optical constants for two different 
Fe/Mg ratios (Mg$_{0.8}$Fe$_{1.2}$SiO$_4$ and Mg$_{1.0}$Fe$_{1.0}$SiO$_4$).
So, although the more Fe-rich olivine fits the data slightly better, 
it only gives a rough estimate of the Fe over Mg ratio for the amorphous 
olivines.

Dorschner et al. (1995) also derived optical constants for amorphous pyroxenes
ranging from MgSiO$_3$ to Mg$_{0.4}$Fe$_{0.6}$SiO$_3$.
Amorphous pyroxenes with the same percentage of Fe as the amorphous 
olivines produce a peak around 21 $\mu$m, which is not seen in our 
ISO-spectrum. Only when the Fe content drops below 0.2 
(i.e. Mg higher than 0.8), the peak vanishes.
For Mg-rich amorphous pyroxenes with a Mg content above 0.8 
only two sets of optical constants are available (x=1 and 0.95 in 
Mg$_x$Fe$_{(1-x)}$SiO$_3$). Since the emissivity of the Fe-poor pyroxenes is 
very dependent on the Fe content, it is difficult to give a reliable 
estimate of the contribution of the amorphous Fe-poor pyroxenes based on these
2 sets only. 
Amorphous pyroxenes, if present at all, will only marginally contribute to 
the IR-excess, due to their lower Fe content compared with the 
amorphous olivines, unless they have an efficient thermal coupling with 
Fe-containing dust, due to an aggregate or a core mantle structure.
Since we have already found indications for separate grain populations, 
see e.g. the temperature
differences found for the amorphous and crystalline olivines,
we assume that these two component grains do not exist and 
only the amorphous Fe-rich olivines are present.

Crystalline forsterite has 3 different crystallographic axes.
We found two studies of the optical constants of the 3 axes, one by
Servoin \& Piriou (1973) who used a synthetic sample and one by Steyer (1974),
who used a natural sample. 
Both data sets do not cover the whole wavelength range.
Although both sets of optical constants should describe the same
material there were quite some differences.
The Steyer (1974) optical constants were not able to reproduce the correct 
wavelengths of the 23.5 and 33.5 $\mu$m bands, found in laboratory 
transmission measurements (Koike et al. 1993; J\"ager et al. 1998), even 
when shape effects or core mantle structures were taken into account.
The Servoin \& Piriou (1973) data set predicts for spherical grains 
peaks at shorter wavelengths than observed. A CDE mixture produces
peaks that are at slightly longer wavelengths than observed. It is likely
that a somewhat modified CDE mixture will give a good match.
The difference between the Steyer and Servoin \& Piriou data may be related to 
the origin of the samples, i.e. respectively natural and synthetic. 
Since Servoin and Piriou only have data longwards of 5 $\mu$m we
extended the data set to the shorter wavelengths using amorphous forsterite
data as published by Scott \& Duley (1996), assuming that the
absorption properties of the amorphous and crystalline grains do not
differ much in this part of the spectrum.
For a fair determination of the abundance of the crystalline forsterite
we will apply Mie theory, since we use the spherical shape also for the other
dust species. However for
the best fit we apply CDE, since this fits the position of the
peaks much better. In this case the abundance is decreased with 25\%
because of the influence of size and shape effects on the peak to continuum
ratio of crystalline forsterite. 

To our knowledge no optical constants are available for the three
different crystallographic axes of the pure Mg-end member of
the crystalline pyroxene solid solution serie, enstatite (MgSiO$_3$).
J\"ager et al. (1998) have measured ortho-enstatite with
a small amount of Fe (Mg$_{0.96}$Fe$_{0.04}$SiO$_3$).
Given the problems mentioned in Sect.~\ref{sec:cryst} 
it is clear that we cannot reproduce 
all the peaks with the desired strength for this set of optical constants,
e.g. the 32.8 $\mu$m peak. 
Therefore we decided not to include enstatite
in our model fit. We will derive the enstatite abundance in 
a different way (see Sect.~\ref{outcome}).

An interesting point is how to combine
the different grain components in model calculations. From an observational 
point of view, there appear to be several separate grain components: the large
temperature difference between the amorphous and crystalline grains
suggest that these components are not in thermal contact, which, 
assuming that they are co-spatial, implies that they should
be treated as two independent grain populations in the radiative transfer.
The situation is less clear for the different crystalline silicates.
They may very well be incorporated in the same grain, and this can change the 
optical properties of the material significantly.
We have tried to produce a forsterite-enstatite-aggregate grain theoretically 
with the Bruggeman (1935) method, and although the results are 
promising we will 
not use these results since no good optical constants for the pure
enstatite are available at the moment.
For forsterite, with its three different crystallographic axes, 
we assumed an equal distribution of the orientation axes. 
This has been achieved by
adopting 3 different grain populations, one for each axis.
The number density, and thus 
the mass, of each population is one third of the total number density (mass) 
of the material. 

Our code is not well suited for large, non-spherical grains.
Therefore for FeO we use instead of Mie a CDE shape distribution, 
which is not strictly correct for the size distribution we expect.
Since shape effects also change the emissivity of FeO, the abundance of FeO
derived from our model fit should be used with great caution.

We have not included melilite, because optical constants with
a sufficient large (0.1-100 $\mu$m) wavelength coverage were not available.

The final point of caution is the fact that we used blackbodies 
to represent the stellar continua.
This will influence the result especially
in the part where absorption from molecules dominates the stellar photosphere,
but not for longer wavelengths where the dust emission dominates.
A more detailed fit of the stellar photospheres will be presented by
van Winckel et al. (in prep.).

\subsection{The best fit model parameters}
\label{outcome}

\begin{table}[ht]
\begin{flushleft}
\begin{tabular}{lrclc}
\hline
\multicolumn{5}{c}{Stellar properties} \\
\hline
L$_{\rm hot\: star}$  &\multicolumn{3}{c}{$1.3 \times 10^{5}$}& L$_{\odot}$ \\
T$_{\rm hot\: star}$  &\multicolumn{3}{c}{$7250 	    $}& K\\
L$_{\rm cool\: star}$ &\multicolumn{3}{c}{$7.4 \times 10^{4}$}& L$_{\odot}$ \\
T$_{\rm cool\: star}$ &\multicolumn{3}{c}{$3750 	$    }	& K\\
\hline
\multicolumn{5}{c}{Shell properties} \\
\hline
Inner radius	& 4.2 &	$\pm $&	$1.0	\times 10^{16}$	& cm \\
Outer radius	& 4.1 &	$\pm $&	$1.0	\times 10^{17}$	& cm \\
$\rho$ at the inner radius & 1.2 & $\pm $&$ 0.2 \times 10^{-18}$& g/cm$^3$ \\
exp in $\rho (R) = C \times R^{\rm exp}$ & -2.2 &$\pm $&$ 0.15$	&  \\
M$_{\rm dust}$	& 3.9 & $\pm$ &$ 1.0	\times 10^{-2}$	& M$_\odot$ \\
$<$\.{M}$>$	& 9.  &	$\pm$ &$ 2.	\times 10^{-4}$	& M$_\odot$/yr \\
\hline
\end{tabular}
\caption{The fixed stellar input parameters together with the Model 
parameters for the best fit, assuming a distance of 3.3 kpc.
The errors for the shell parameters are internal model errors. }
\label{shellfit}
\end{flushleft}
\end{table}

\begin{figure*}[ht]
\centerline{\psfig{figure=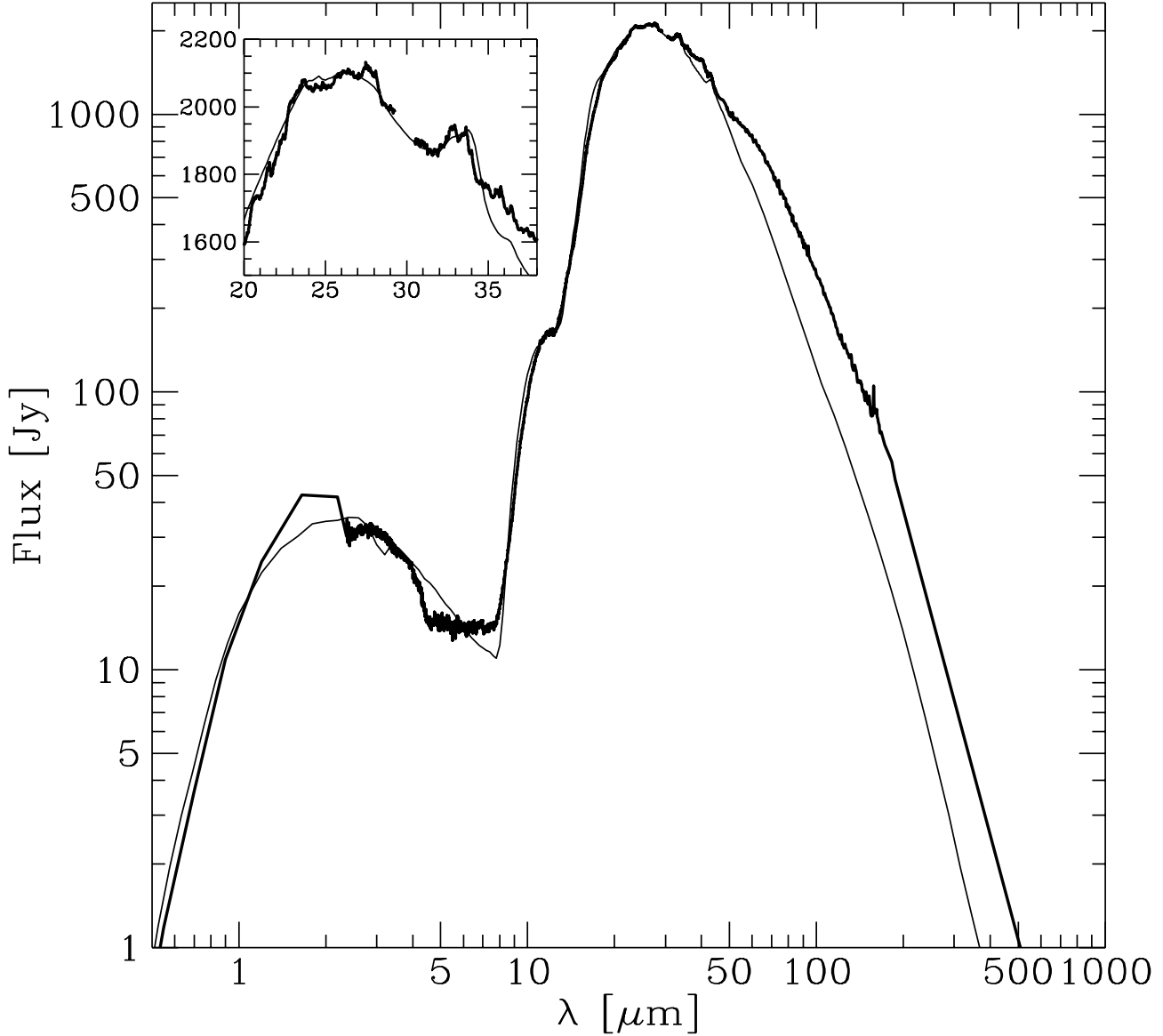,width=180mm}}
\caption[]{The SED of AFGL 4106 (thick line) and our best model fit (thin line)
The input values are those from Table~\ref{tab:dustfit}, with the exception
that enstatite was not put in and for forsterite and FeO we applied CDE instead
of Mie.} 
\label{fig:dustfit}
\end{figure*}
Our best fit to the SED can be seen in Fig.~\ref{fig:dustfit}. 
The model parameters of the shell are given in Table~\ref{shellfit}, 
together with an estimate of their errors.
Starting estimates for the inner and outer radius of the dust shell were
derived from the TIMMI image, assuming a distance to AFGL~4106 of 3.3 kpc.
\begin{table}[ht]
\begin{flushleft}
\begin{tabular}{lc}
\hline
\multicolumn{2}{c}{Dust properties} \\
\hline
Grain size ($\mu$m)					&  0.4 - 6.0 \\
Mass fraction of crystalline forsterite (Mg$_2$SiO$_4$) &  4\%	\\
Mass fraction of crystalline enstatite (MgSiO$_3$) 	&  4-11\%	\\
Mass fraction of crystalline water ice (H$_2$O)		&  5\%	 \\
Mass fraction of crystalline corundum ($\alpha$-Al$_2$O$_3$) 	&  17-15\% \\
Mass fraction of amorphous olivine (Mg$_{0.8}$Fe$_{1.2}$SiO$_4$)&  70-65\% \\
Mass fraction of ironoxide (FeO)				&  $< 1$\% \\
\hline
\end{tabular}
\caption[]{The dust properties as derived from the radiative transfer 
model fitting. The ranges in the mass fractions come from the 
uncertainty about the enstatite to forsterite ratio.}
\label{tab:dustfit}
\end{flushleft}
\end{table}
This value will be discussed in Sect.~\ref{distance}.
The dust shell is optically thin at most wavelengths.
This implies that the inner and outer radius of the dust shell 
depend linearly on the distance, as does the duration of the mass-loss burst.
The total mass in the shell scales roughly with the square of the distance,
therefore the mass loss rate will also scale linearly
while the density at the inner radius is inversely proportional to 
the distance.

In Table~\ref{tab:dustfit} the dust parameters used in the model are given.
To minimize the number of free parameters, we have assumed that all
grain-species have the same size distribution.
The typical errors in the grain size are 25\%.
In all models we have assumed a size distribution
$N(a) \propto a^{-3.5}$ (Mathis et al. 1977).
 
Apart from the uncertainty of the enstatite mass fraction, which influences
the whole scheme, the errors in the mass fractions differ between the 
materials. For FeO the error is difficult to determine since
the shape distribution largely 
influences the outcome, an order of magnitude cannot be excluded.
The mass fraction of the amorphous olivines is accurate
to a few percent of the total mass.
However the chemical Fe:Mg ratio is not well determined in the amorphous
olivines, a change of 10\% in this ratio will only marginally influence the 
spectrum. 
The crystalline forsterite abundance is mainly based on the strength
of the 23.6 and 33.6 micron peak and has an error of about 1\% of the total 
mass.
The abundance of enstatite
is hard to determine since no optical constants were available for this 
material. So, we used an alternative way to determine the enstatite to 
forsterite ratio.
From transmission measurements of enstatite and forsterite one may determine
the wavelength dependent extinction coefficients.
The 33.6 $\mu$m peak is caused by forsterite and the 32.8 $\mu$m
peak by ortho-enstatite. So the peak-strength ratio of these two features 
is indicative for the abundance ratio. Since the peak wavelengths of these two
features are very close, temperature effects do not influence the results.
If we use the transmission spectra published by Koike \& Shibai, we find a 
ratio of enstatite to forsterite of 3. However, transmission spectra of 
forsterite published by J\"ager et al. (1998) give a peak strength of the
33.6~$\mu$m feature which is 3 times smaller than that of 
Koike \& Shibai (1998), from which a ratio of enstatite to forsterite of 1 
follows. We conclude that, given the large spread in laboratory data
published so far, we find an abundance ratio of enstatite to forsterite
between 1 and 3. 
It would be interesting to compare this value with the olivine to pyroxene 
ratio seen for the amorphous silicates.
However, because of reasons given in Sect.~\ref{input}
this is not possible at the moment.

Since we were not able to reproduce (in a single dust shell) the low 
temperature of crystalline H$_2$O ice, the abundance of crystalline H$_2$O ice 
is only based on the strength of the 43$\mu$m peak and is in that sense 
accurate to about 1\% of the total mass.
However, if there is a very cool H$_2$O ice dust shell surrounding
this object, which can be responsible for the LWS excess and missing
strength of the 60 $\mu$m feature, 
the H$_2$O ice abundance would probably increase at least an order of 
magnitude.

We tried different models and only with large grains we were able to
fit the shape, strength and position of the 10 $\mu$m feature. 
Smaller grain sizes (spherical or not) would produce a distinct
and relatively sharp peak at shorter wavelengths in contrast to the flat 
spectrum we observed.
The presence of these large
grains causes a grey extinction at optical wavelengths. 
This influences our previous
determination of the extinction (see Sect.~\ref{extinction})
and therefore the luminosity of both stars. 
The luminosity of the stars was derived using
the standard interstellar extinction law (Fluks et al. 1994) 
for both the interstellar and circumstellar extinction.
Since the colour excess changes much less when dealing with large grains,
the stars are brighter than derived earlier in this paper.
Using an iterative fitting scheme we re-derived the luminosity and
interstellar extinction.
We find a luminosity of $1.2 \times 10^4 \, d^2 $L$_\odot$
for the hot star and $6.8  \times 10^3 \, d^2 $L$_\odot$ for the
cool star and an interstellar extinction of $E(B-V) = 1.1$ mag.
Note that this value for the interstellar extinction is very close to the
value which has been derived from the DIBs.
Furthermore, our model predicts an optical depth to the star of
$\tau = 0.6$ at 6583 \AA, which is in very good agreement
with the value $\tau = 0.6 \pm 0.1$ found by van Loon et al. (in prep.) for
the optical depth at [N {\small II}]-line.
Although $\tau$ at 6583 \AA\ is 0.6, the circumstellar $E(B-V)$ is 
only 0.015 mag., 
due to the grey extinction (at optical wavelengths) of the large grains.  
So, the problem between the total extinction on one side and 
the circumstellar and interstellar dust 
extinction on the other side
is resolved by the presence of large grains in the circumstellar
environment leading to a non-standard extinction law for the dust shell.

Even with these large grains we were not able to fit the entire
shape of the 10 $\mu$m silicate feature. Especially around 13 $\mu$m our models
predicted lower flux levels than observed. When we subtract our model fit
from the observations, a 13 $\mu$m feature is evident.
This feature resembles the 13 $\mu$m feature observed in AGB stars with
low mass loss rates, and is often attributed to corundum
($\alpha$-Al$_2$O$_3$) (Onaka et al. 1989; Begemann et al. 1997;
Kozasa \& Sogawa 1998).
Therefore we also included corundum in our modelling, which indeed gives 
a much better fit around 13 $\mu$m.

The total mass in the shell, assuming a gas-to-dust ratio of 100 and a 
distance of 3.3 kpc, 
is 3.9 M$_\odot$. Taking all the different errors and
assumptions (such as the spherical shape of the dust grains) into account, 
this mass may be overestimated by up to a factor two.

While about 70~\% of the dust shell consists of Fe-bearing amorphous 
olivines,
about 7 to 15~\%, depending on the assumed enstatite to forsterite ratio,
constitutes of Mg-rich crystalline silicates. 
For the crystalline silicates only the end members of the olivine 
and pyroxene series, forsterite and enstatite respectively, are present.
We note that the crystalline silicates only become important at the longer 
wavelengths. At 10 $\mu$m the amorphous silicates show a prominent peak
but in this wavelength region there is no sign of the presence of the
crystalline silicates.
We conclude that the temperature of the crystalline silicates 
is lower than that of the amorphous ones. This temperature difference
is confirmed by our model calculations. The temperatures at the inner radius
are roughly 160, 125 and 100~K, for respectively the amorphous olivines,
crystalline forsterite and the crystalline water-ice.
These temperatures are higher than estimated in 
Sect.~\ref{sec:ident}, but one should take into account that there is a 
temperature gradient and that the values mentioned here are maxima.
The temperature difference is caused
by the increased opacity in the UV, optical and near-IR 
when Fe is incorporated in the silicate.
This implies that the two species are not 
thermally coupled, suggesting separate grain populations.

\subsection{Discussion of the model fit}

For some parts of the spectrum we did not obtain a satisfactory fit.
The most obvious part is the LWS spectrum.
We were not able to fit simultaneously the slope in the LWS part and
the peak in the dust emission in the SWS part. 
A shallower density gradient would result in a broadening of the synthetic
spectrum around 30 $\mu$m.
We decided to fit the peak instead of the long wavelength part, also 
because of the differences between the 100 $\mu$m IRAS and LWS flux in this 
region, which required the steep density gradient.
Note that the model spectrum convolved with the IRAS 100 $\mu$m sensitivity
gives a flux of 125 Jy, which is lower than both the IRAS and LWS flux.
The fact that the flux in our model is too low and that the LWS spectrum 
contains a prominent and very cool H$_2$O ice band, gives an 
indication that an extra cool dust component, with likely a large
amount of crystalline H$_2$O-ice, surrounding this object is required
in the LWS beam. The presence of a large amount of crystalline H$_2$O, which
is generally not found in the ISM, suggests that this extra contribution 
is due to a previous mass loss phase. 
The discrepancy between IRAS and ISO, however, remains unsolved.
GKH had a 3 sigma detection of AFGL~4106 at 1.3 mm of $27 \pm 9$~mJy.
In our model we find a flux of 17 mJy; this is within the errorbars.
The ratio of 1.3~mm to (IRAS) 100 $\mu$m flux densities is reproduced by
our model, indicating that the largest grains are significantly smaller than 
100 $\mu$m.

Differences between the model and the broadband photometry
in the optical and near-infrared, where we see the photospheres
of both stars, can probably all be explained by
the fact that we used blackbodies instead of real stellar photospheres.
A third discrepancy is between 4.5 and 8 $\mu$m.
We were not able to fit this flat part of the spectrum.
Part of it can be explained by photospheric CO absorption (around 4.5 $\mu$m).
A few experiments with FeO seem to indicate that a proper treatment of the
shape effects of FeO might produce a flat spectrum in this area.
On the other hand, the TIMMI N-band image shows a bright central emission, 
which cannot only be explained by the photospheres of the stars.
It seems that an extra (hot) dust component is present at the
center and that its infrared excess extends to the shorter wavelengths.
Whether this represents a recent mass loss phase or a more stable
circumstellar dust configuration (circumbinary disk) is unknown. 

Previous modelling of AFGL~4106 has been performed by Volk and Kwok (1989) and
HKV, who presented the same model, and GKH.
Both models use in the center a single blackbody, with a temperature of 
5000 (HKV) and 4750~K (GKH),
surrounded by a single dust component circumstellar envelope.
Because of these big differences between their and our model
it is not useful to compare the inner radius and
density at the inner radius found by them and us.

It is interesting that HKV and the present study both find a density 
distribution which differs from $\rho (R) \propto  R^{-2.}$,
with $R$ the distance of the position in the dust shell
to the central object.
HKV found a steeper density gradient than we did, but this is within the
error margins.
The main difference between our parameters and the ones of GKH is the exponent
in the density gradient. Since we have now spectral coverage longwards
of 25~$\mu$m, we can exclude the possibility of a density gradient of 
$\rho \propto R^{-1}$. Their fit predicts a much broader IR excess 
around 25 $\mu$m than observed. 
It should be noted that GKH fit the IRAS 60 and 100 $\mu$m points
better than the present study. However, because of the significant 
broadening of the top 
of the energy distribution when applying a flatter density distribution,
which is not observed,
the IR-excess problem is more likely solved with an extra dust shell.
The density gradient of $\rho(R) \sim R^{-2.2}$ can be caused by 
two reasons:
(I) The mass loss rate increased or (II) the outflow velocity decreased
with time. In both cases the density gradient increases
due to the geometric expansion of the dust shell with time.
So whatever the reason is for the
present steep gradient, it is intrinsic but has steepened during the expansion
of the dust shell.

Assuming different grain size distribution for the different dust
species might lead to an improvement of the fit and will result
different mass ratios for the dust species. A smaller size distribution
for one dust component will result in a decrease of its mass ratio and 
vice versa.

\section{Discussion}
\label{discussion}

\subsection{The distance and luminosity}
\label{distance}

In the literature 
distances ranging from 0.67 -- 3.3 kpc (Garc\'{\i}a-Lario et al. 1994) are 
given. 
However one should note that AFGL 4106 is located in the direction of Carina,
so we are looking along a spiral arm, which makes it hard to 
relate column density and distance.
Fortunately, we have spatial information from the N-band image. 
The inner radius of the dust shell is about
$1.^{\prime\prime}$, and together with
the expansion velocity found in CO (Josselin et al. 1998) and 
[N{\small II}]-line (van Loon et al., in prep.) of 30 to 40 km/s and the 
present-day shape of the SED, one can put a lower limit to
its distance. Assuming that the outflow is spherical, which seems plausible 
according to our modelling results and the [N {\small II}]-line
velocity distribution by van Loon et al. (in prep.), and that the velocity
has not changed significantly in time, one can derive a kinematical
age for the dust shell of 135 years times the distance in kpc.
Due to the expansion of the dust shell
it is expected that the circumstellar dust extinction becomes smaller, 
leading to an increase in the brightness at short wavelengths.
AFGL~4106 has already been identified in the
Cape Observatory Photographic Durchmusterung from 1895 to 1900 
(Gill and Kapteyn 1900) with a photographic magnitude $M_{\rm pg} = 9.9$ mag.
In the Cordoba Durchmusterung (Thome 1914) a visual magnitude of
9.4 has been estimated for AFGL~4106.
In the extended Henry Draper Catalogue, with 
observations taken between 1922 and 1937 (Cannon and Mayall 1949), 
its photographic 
magnitude was $10.0 \pm 0.3$. On 18 March 1988 Hrivnak et al. (1989)
measured a B magnitude of $10.24 \pm 0.01$ 
($M_{\rm pg} = M_{\rm B} - 0.11$) and a V magnitude of 
$8.73 \pm 0.01$. 
From these observations one can derive that
$\Delta M_{\rm pg} = 0.2 \pm 0.5$ mag. over the last 90
years and $\Delta M_{\rm V} = -0.7 \pm 0.5$ mag. over the last 75 years,
assuming an error of 0.5 magnitude in both Durchmusterungen.
So the visual and photographic band trends contradict each other, 
although the errorbars are significant. 
Since the hot star is the dominant component at these two wavelengths
it is possible that the star lowered its temperature, from a temperature
with its peak luminosity at the photographic band to a temperature with its 
peak luminosity at the visual band.
However, this is in contradiction with the present temperature, which peaks
around the photographic band and is also not expected from evolutionary
considerations. 

In any case, the stability of the optical brightness of AFGL~4106 over a 
timescale of $\approx 90$ years can be used to constrain its distance and 
luminosity.
Using our dust model fit we can calculate the expected SED of the system and 
the visual and photographic brightness in the past,
assuming that the dust composition, the outflow velocity, the temperature
and the radii of the stars did not change.
For the photographic magnitude, it is safe to assume
that the star has not brightened more than 0.3 magnitude over the last
90 years. This puts the star at a distance of at least 3.15 kpc and probably
further away.
On the other hand the brightning in the $V$-band over the last 75 years
of $\Delta M_{\rm V} = -0.7 \pm 0.5$ puts the star at a distance of
$1.5^{+2.1}_{-0.4} $ kpc.
There is only a very small overlap region between these two trends, 
suggesting a distance of 3.3 kpc, just at the upper limit quoted by
Garc\'{\i}a-Lario et al. (1994).
The size of this overlap region is mainly based on
the estimated errors of the Cape Observatory Photographic 
Durchmusterung and the Cordoba Durchmusterung.

This distance of 3.3 kpc implies that the luminosity is $1.3 \times 10^5$ and 
$7.4 \times 10^4$ L$_{\odot}$ for the hot and cool star respectively,
again pointing to a massive binary system.
Their luminosities point to stars with a main sequence mass between 15 and 20
M$_{\odot}$, using evolutionary tracks by Maeder \& Meynet (1988).
Other properties of AFGL~4106 also support the high masses of both stars.
The expansion velocity of 30 to 40 km/sec is unusually high for AGB stars, and 
the relative large grains are typically found in the outflows massive 
supergiants (Jura 1996).

It is likely that the warm star is responsible for the expelled
dust shell, and is now in the very rare evolutionary phase of the 
post-Red-Supergiants, which makes it a blue supergiant or WR star progenitor. 
Only a few other stars are in the same evolutionary status, e.g. IRC+10420
(Jones et al. 1993), and the central star of the radio nebula G79.29+0.46
(Trams et al. 1999).

\subsection{The dust shell}
\label{dust_evol}

At a distance of 3.3 kpc the mass loss episode took about $4.3\times 10^3$
years and stopped 450 years ago. During this period the star
expelled $3.9 $ M$_{\odot}$, assuming a gas to dust ratio of 100.
This implies that the average mass loss rate is
roughly $9.\times 10^{-4}$~M$_{\odot}$/yr.

In our modelling we assumed spherical symmetry, but we do see
substructure in our N-band images. This is also
expected, since it is likely that the M-type supergiant will influence the 
spatial distribution of the dust and gas, expelled by the hot star,
by transferring orbital momentum to the gas and dust.
This effect can create a density enhancement in the plane of the binary.
This possible enhancement can lead to an over- or under-estimate of the 
total mass, depending on the angle at which we observe this system.
However, the shape of the [N {\small II}]-line derived from different 
rotation angles of the slit, suggest that the shape of the dust shell is 
close to a sphere (van Loon et al., in prep.).
This implies that the orbital seperation between the two stars
is probably relatively large and therefore the influence of the companion 
on the mass-loss not so important. 
So, although the spherical symmetry and other assumptions used in our model 
calculations might be somewhat simple, it is not expected that the 
mass-loss rate and the total mass found in our calculations would change more 
than a factor 2.

\subsection{Formation of crystalline silicates}
\label{sec:crystal}

The formation of crystalline silicates is still largely unknown.
Several researchers have tried to quantify the condensation sequence of 
oxygen rich dust (e.g. Tielens, 1990; Gail, 1998). 
However, these are mainly based on the different condensation temperatures,
therefore assuming thermal equilibrium; it is not known if
this assumption is valid.
Condensation of dust species and chemical reactions will only occur if:
(I) the density is high enough and (II) the temperature is suitable (low enough
for condensation and high enough for chemical reactions).
Since both the density and the temperature decrease with the distance from the
star, it is likely that at a certain moment the dust structure freezes out.
It is this stable dust configuration which we see at the moment.
Taking the condensation sequences mentioned above,
it is expected that the first silicate that will form is forsterite. 
When the temperature 
becomes lower and the density is still high enough this will transform into
enstatite. Both species are expected to form above the glass
temperature and are therefore thought to be crystalline.
We see evidence in our spectrum for both forsterite and enstatite.
This implies that the forsterite to enstatite transition is stopped
before the forsterite was completely transformed.
There might be 3 reasons for this: (I) The density became too low, (II)
the temperature dropped too rapidly or (III) other chemical reactions
took over. It is likely that it is a combination of these three.
One chemical reaction that could take place at lower temperatures
is the incorporation of Fe into the silicates. Since our spectra show that
the crystalline silicates are very Mg- rich and the amorphous silicates
contain a lot of Fe, it is likely that this incorporation of Fe
results in a destruction of the crystal structure.
The clear chemical separation between the amorphous and crystalline materials
is intriguing. It appears that the inclusion of Fe in Mg-rich silicates
is a kind of runaway process. 

Tielens et al. (1998) proposed a scenario for such a chemical separation
between amorphous and crystalline silicates.
When the temperature becomes low enough Fe may react with the 
Mg-rich crystalline silicates. The opacity will increase, due to the
incorporation of Fe, and therefore the temperature of the grain.
This process will act as a thermostat, incorporating just sufficient Fe
in the grains to keep the temperature near 800~K where Fe can just diffuse in.
This temperature is below the glass temperature for Fe-bearing silicates and
the lattice cannot attain its energetically most favorable structure, thus 
leading to an amorphous structure.
Because of the increased opacity due to the incorporation of Fe,
it is likely that the temperature of the other grains behind this
Fe-reaction zone will decrease, leading to an even higher difference between 
the Fe-rich (amorphous) and Mg-rich (crystalline) silicates if
thermal coupling between gas and dust can be neglected.
Grains that already contain a small amount of Fe will be hotter and therefore 
the reaction rate for the adsorption of Fe will be higher then for grains 
without any Fe.
This would lead to the required runaway process.
If this scenario is correct, it would imply that the crystalline
silicates are the primordial condensation products, and this will give us 
new insights in the dust nucleation.
Although this would lead to the required chemical separation between the 
crystalline and amorphous materials, it is not clear why the amorphous olivines
should contain more Fe than the amorphous pyroxenes, if present at all.

Besides the forsterite-enstatite condensation sequence, Tielens~(1990) 
also presented a condensation sequence starting with
corundum. The next condensation product would then be melilite, 
for which we find some evidence.
According to the same scheme diopside
would form. Since certain strong features of this material are not found
in our ISO-spectrum, this may imply that the dust forming process along this
condensation sequence line froze out around amorphous melilite.

Gail and Sedlmayr (submitted to A\&A) applied non-equilibrium calculation for
the dust formation in outflows of M stars. They do explain the 
formation of several dust species, such as the amorphous Fe-rich olivines,
however they do not consider, based on laboratorium experiments, 
the formation of pyroxenes and are not capable to explain the
co-existence of pure forserite particles and Fe-rich
amorphous olivines.

\section{Conclusions}
\label{conclusion}

The main results of this study can be summarized as follows:
\begin{itemize}
\item[1] AFGL~4106 is a (double-lined spectroscopic) binary, 
consisting of a warm star with $T_{\rm eff} =7250$~K and a cool companion with 
$T_{\rm eff} = 3750$~K.
\item[2] The luminosity ratio of both stars 
L$_{\rm warm}$/L$_{\rm cool} = 1.8$, which indicates that both stars are 
evolved. The warm star is likely responsible for the dust shell.
\item[3] The expansion velocity of the shell, the grain size distribution and 
the minimal variations in the photographic and V band over the last 90
years all suggest a high luminosity. We conclude that the stars are at 
$3.3  \pm 1.0$ kpc and have L/L$_\odot = 1.3 \times 10^5$ and $7.4 \times 10^4$
respectively.
These values correspond to main sequence masses of 15 to 20 M$_\odot$.
The warm star is probably evolving rapidly to the blue part of
the HR-diagram and may evolve to a blue-supergiant (ending its
life like SN 1987A) or to a Wolf-Rayet phase.
\item[4] The main component of the dust are large ($\approx 1 \mu$m) 
amorphous, Fe-rich olivines. If also amorphous pyroxenes are present,
they will have a much lower Fe content.
We find a rich spectrum of narrow solid state emission bands
in the ISO-SWS and LWS spectra, which we identify with
crystalline olivines and pyroxenes. These grains are Mg-rich and Fe-poor
and have an abundance between 7 and 15 \% by mass, depending on the assumed 
enstatite to forsterite ratio. 
We also find evidence for the presence of FeO, Al$_2$O$_3$, melilite and 
crystalline H$_2$O-ice. Shape effects can have an important influence
on the derived abundances. 
\item[5] The temperatures for the different dust species are quite different, 
this is caused by variations in their absorption characteristics at visual 
and near-IR wavelengths. These temperature differences 
imply that the different dust species are not thermally coupled, 
directly (e.g. as a composite grain) nor indirectly (e.g. via gas-dust 
interactions).
\item[6] The crystalline silicate bands can be divided into narrow
(FWHM/$\lambda < 0.03$) and broad (FWHM/$\lambda > 0.03$) features.
The wavelengths of the broad features match well with forsterite, while
those of most of the narrow features line up well with enstatite.
The width of the features is significantly smaller than those
observed in laboratory spectra.
The abundance ratio of enstatite to forsterite is between 3 and 1.
\item[7] The mean mass-loss rate was $\approx 9. \times 10^{-4}$ M$_\odot$/yr
for a period of about $4.3\times10^3$ years and probably increased 
during this period. The mass loss stopped 450 years ago.
The total mass expelled, assuming a gas to dust ratio of 100, is 
3.9~M$_\odot$.
\item[8] We find a cool dust component, which cannot be fitted by
our dust model. This component is either an older mass loss phase, or
an incorrect background subtraction of the ISM.
We also found indications from the spectrum between 4 and 7 $\mu$m
and the 10 $\mu$m imaging that a 
third dust component, close to the binary is present.
Whether this represents a recent mass loss phase or a more stable
circumstellar dust configuration is not known. 
\end{itemize}
Because of its (IR-)brightness AFGL~4106 will be a key object in the further
study of (crystalline) dust formation. 
High resolution imaging and spectroscopy both in the optical and IR 
will allow us 
to constrain the circumstellar versus interstellar extinction, the photospheres
of the components, density and abundance gradients through the CSE, and thus 
the conditions for dust formation and evolution.

\acknowledgements{The authors wish to thank the SWS
Instrument Dedicated Team, in particular Th. de Graauw, 
for their support in obtaining the spectra. We thank A.G.G.M. Tielens
for fruitful discussion.
FJM acknowledges support from NWO grant 781-71-052. 
LBFMW, AdK and JB acknowledge
financial support from the Royal Netherlands Academy of Arts and
Sciences, and from an NWO 'Pionier' grant. }


\begin{thebibliography}{}

\bibitem[]{}
Barlow M.J., 1998, A\&SS 255, 315
\bibitem[]{}
Begemann B., Dorschner J., Henning Th., et al., 1997, ApJ 476, 199 
\bibitem[]{}
Bertie J.E., Labb\'{e} H.J., Whalley E., 1969, The Journal of Chemical 
Physics 50, 4501
\bibitem[]{}
Bohren C.F., Huffman D.R., 1983, Absorption and scattering of light by 
small particles., John Wiley and Sons, Inc., New York
\bibitem[]{}
Bouwman J., Waters L.B.F.M., 1998, A\&SS 255, 435
\bibitem[]{}
Bruggeman D.A.G., 1935, Ann. Physics (Leipzig) 24, 636 
\bibitem[]{}
Cannon A.J., Mayall M.W., 1949, Ann. Astron. Obs. Harvard Coll. 112, 1
\bibitem[]{}
Clegg P.E., Ade P.A.R., Armand C., et al., 1996, A\&A 315, L38
\bibitem[]{}
Decin L., Cohen M., Eriksson K., et al., 1997, in Proceedings of the First 
ISO Workshop on Analytical Spectroscopy with SWS, LWS, PHT-S and CAM-CVF,
ed. Heras A.M., Leech K., Trams N.R., Perry M., ESA Publications 
Division, Noordwijk, The Netherlands
\bibitem[]{}
Dorschner J., Begemann B., Henning Th., et al., 1995, A\&A 300, 503
\bibitem[]{}
Ehrenfreund P., Cami J., Dartois E, Foing B.H., 1997, A\&A 317, L28  
\bibitem[]{}
Fluks M.A., Plez B., The P.S., et al., 1994, A+AS 105, 311 
\bibitem[]{}
Gail H.-P., 1998, A\&A 332, 1099
\bibitem[]{}
Garc\'{\i}a-Lario P., Manchado A., Parthasarathy M., Pottasch S.R., 1994, A\&A 285, 179
\bibitem[]{}
Garc\'{\i}a-Lario P., Manchado A., Pych W., Pottasch S.R., 1997, A\&AS 126, 479
\bibitem[]{}
Gill D., Kapteyn J.C., 1900, Ann. Cape Obs. Edinburgh, vol. V 
\bibitem[]{}
de Graauw Th., Haser L.N., Beintema D.A., et al., 1996, A\&A 315, L49
\bibitem[]{}
G\"{u}rtler J., K\"{o}mpe C., Henning Th., 1996, A\&A 305, 878
\bibitem[]{}
Henning Th., Mutschke H., 1997, A\&A 327,743
\bibitem[]{}
Henning Th., Begemann B., Mutschke H., Dorschner J., 1995, A\&AS 112, 143
\bibitem[]{}
Herbig G.H., 1995, ARA\&A 33, 19
\bibitem[]{}
Hrivnak B.J., Kwok S., Volk K.M., 1989, ApJ 346, 265
\bibitem[]{}
IRAS explanatory supplement 1988, eds. C. Beichmann, et al. 
\bibitem{kay}
J\"ager C., Molster F.J., Dorschner J., et al., 1998, A\&A 339, 904
\bibitem[]{}
Jenniskens P., D\'esert F.-X., 1994, A\&ASS 106, 39
\bibitem[]{}
Jones T.J., Humphreys R.M., Gehrz R.D., et al., 1993, ApJ 411, 323
\bibitem[]{}
Josselin E., Loup C., Omont A., et al., 1998, A\&AS 129, 45
\bibitem[]{}
Jura M., 1996, ApJ 472, 806
\bibitem[]{}
K\"{a}ufl H-.U., Jouan R., Lagage P.O., et al., 1994, Infrared Phys. 
Technol. 35, 203
\bibitem[]{}
Kessler M.F., Steinz J.A., Anderegg M.E., et al., 1996, A\&A 315, L27
\bibitem[]{}
Koike C., Shibai H., 1998, ISAS report no. 671
\bibitem[]{}
Koike C., Shibai H., Tuchiyama A., 1993, MNRAS 264, 654
\bibitem[]{}
Koike C., Kaito C., Yamamoto T., et al., 1995, Icarus 114, 203
\bibitem[]{}
Kozasa T., Sogawa H., 1998, A\&SS 255, 437
\bibitem[]{}
Kurucz R.L., 1991, in Precision Photometry: Astrophysics of the Galaxy,
Ed. Davis Philip A.G., Upgren A.R., Janes K.A., L. Davis Press, Schenectady 
\bibitem[]{}
Little-Marenin I.R., Little S.J., 1988, ApJ 333, 305 
\bibitem[]{}
Little-Marenin I.R., Little S.J., 1990, AJ 99, 1173
\bibitem[]{}
Lucy L.B., 1974, AJ 79, 745
\bibitem[]{}
Maeder A., Meynet G., 1988, A\&AS 76, 411
\bibitem[]{}
Mathis J.S., Rumpl W., Nordsieck K.H., 1977, ApJ 217, 425
\bibitem[]{}
Monnier J.D., Geballe T.R., Danchi W.C., 1998, ApJ 502, 833
\bibitem[]{}
Mutschke H., Begemann B., Dorschner J., et al., ,1998, A\&A 333, 188 
\bibitem[]{}
Omont A., Forveille T., Moseley S.H., et al., 1990, ApJ 355, L27
\bibitem[]{}
Onaka T., de Jong T., Willems F.J., 1989, A\&A 218, 169
\bibitem[]{}
Schaeidt S.G., Morris P.W., Salama A., et al.: 1996, A\&A 315, L55
\bibitem[]{}
Scott A., Duley W.W., 1996, ApJS 105, 401 
\bibitem[]{}
Servoin J.L., Piriou B., 1973, Phys. Stat. Sol. (b) 55, 677
\bibitem[]{}
Steyer T.R., 1974, PHD-thesis, University of Arizona
\bibitem[]{}
Swinyard B.M., Clegg P.E., Ade P.A.R., et al., 1996, A\&A 315, L43
\bibitem[]{}
Tielens A.G.G.M., 1990, in From Mira's to Planetary Nebulae: Which
path for stellar evolution?, eds. M.O. Mennessier \& A. Omont, 186
\bibitem[]{}
Tielens A.G.G.M., Waters L.B.F.M., Molster F.J., Justtanont K., 1998,
A\&SS 255, 415
\bibitem[]{}
Thome J.M., 1914, Results of National Argentine Obs. 21, 1
\bibitem[]{}
Trams N.R., Van Tuyll C.I., Voors R.H.M., et al., 1999 
in IAU Colloquium 169, Variable and Non-spherical stellar winds 
in luminous hot stars", Eds. Wolf, Fullerton, Stahl
\bibitem[]{}
Valentijn E.A., Feuchtgruber H., Kester D.J.M., et al., 1996, A\&A 315, L60
\bibitem[]{}
Volk K.M., Kwok S., 1989, ApJ 342, 345
\bibitem[]{}
Waters L.B.F.M., Molster F.J., de Jong T., et al., 1996, A\&A 315, L361
\bibitem[]{}
Waters L.B.F.M., Waelkens C., van Winckel H., et al., 1998, Nat 391, 868

\end{thebibliography}
\end{document}